\documentclass[a4paper, onecolumn, unpublished, noarxiv, 10pt]{quantumoriginal}

\usepackage{graphicx}
\usepackage{dcolumn}
\usepackage{bm}
\usepackage[dvipsnames, table]{xcolor}
\usepackage{braket}
\usepackage{svg}
\usepackage{quantikz}
\usepackage{yquant}
\usepackage{amsmath}
\usepackage{amssymb}
\usepackage{stmaryrd}
\usepackage{bbold}
\usepackage{mdframed}
\usepackage{multirow}
\usepackage{tabularx}
\usepackage{enumitem}
\usepackage{caption}
\usepackage{subcaption}
\usepackage[most]{tcolorbox}
\usepackage{hyperref}
\DeclareMathOperator{\tr}{tr}
\DeclareMathOperator{\Tr}{Tr}

\newtcolorbox{boxenv}[1]{colback=white, colframe=black, fonttitle=\bfseries, title=#1}
\newlength{\slength}
\makeatletter
\newcommand{\myfirstpagefootnote}[1]{%
  \insert\footins{%
    \footnotesize\sffamily
    \iftoggle{@titlepage}
      {\hsize\dimexpr\linewidth\relax\@parboxrestore}%
      {\ifbool{@twocolumn}
        {\hsize\dimexpr0.5\linewidth-0.5\columnsep\relax\@parboxrestore}%
        {\hsize\dimexpr\linewidth\relax\@parboxrestore}%
      }%
    #1%
  }%
}
\apptocmd{\@printauthors}{%
  \myfirstpagefootnote{GitHub repository: \url{https://github.com/Quantinuum/SyQMA}}%
}{}{}
\makeatother

\begin{document}

\title{SyQMA: A memory-efficient, symbolic and exact universal simulator for quantum error correction}
\author{George Umbrarescu}
\affiliation{Quantinuum, Partnership House, Carlisle Place, London SW1P 1BX, United Kingdom}
\affiliation{Department of Physics and Astronomy, University College London, London WC1E 6BT, United Kingdom}
\orcid{0000-0003-4376-2147}
\email{george.umbrarescu.20@ucl.ac.uk}
\author{David Amaro}
\affiliation{Quantinuum, Partnership House, Carlisle Place, London SW1P 1BX, United Kingdom}
\orcid{0000-0001-7853-9581}
\email{david.amaro@quantinuum.com}

\begin{abstract}
The classical simulation of universal quantum circuits is crucial both fundamentally and practically for quantum computation. We propose SyQMA, a simulator with several convenient features, particularly suited for quantum error correction (QEC). SyQMA simulates universal quantum circuits with incoherent Pauli noise and computes exact expectation values and measurement probabilities as symbolic functions of circuit parameters: rotation angles, measurement outcomes, and noise rates. This simulator can sample measurement outcomes, enabling the simulation of dynamic quantum programs where circuit composition depends on prior measurement outputs. For QEC, it performs circuit-level maximum-likelihood decoding, provides exact symbolic expressions for logical error rates, and verifies the fault distance of fault-tolerant (FT) stabiliser and magic state preparation protocols. These features are enabled by an intuitive extension of stabiliser simulators, where each non-Clifford Pauli rotation and incoherent Pauli channel is compactly represented via auxiliary qubits and a modified trace. 
Representing the state requires only polynomial memory and time, while computing expectation values and measurement probabilities takes exponential time in the number of non-Clifford rotations and deterministic measurements, but only polynomial memory.
The FT preparation of stabiliser and magic states, including the first stage of magic state cultivation, is analysed without approximations.
We also exactly convert the disjoint error probabilities of a general multi-qubit Pauli channel to independent ones, a key step for creating and sampling from detector error models.
The code is publicly available and open-source.
\end{abstract}

\maketitle

\section{Introduction}

Quantum computing has the potential to solve many difficult problems in fields such as chemistry, logistics and financial services. While access to quantum computers is limited, researchers rely on simulators to verify and debug the execution of quantum programs. In particular, simulators and emulators running on classical computers are essential to anticipate the performance of quantum algorithms on real, noisy quantum hardware. 

Quantum error correction (QEC) stands to benefit especially from such capabilities: beyond debugging, simulation is required to predict logical performance and guide the design of fault-tolerant protocols under realistic noise. While many QEC programs are Clifford and efficiently simulable via the stabiliser formalism, practical fault-tolerant computing increasingly involves non-Clifford resources, for instance in magic state distillation \cite{bravyi2005universal} and cultivation \cite{gidney2024magic}, and demands precise characterisation of logical error rates (LERs) that are sometimes too low for Monte Carlo methods.

Meeting these QEC-driven requirements simultaneously remains challenging for existing tools. Exact, strong simulation of universal circuits is generally exponentially costly, and the difficulty is compounded when one requires (i) exact (non-stochastic) treatment of circuit-level noise,
(ii) symbolic access to circuit parameters (noise rates, rotation angles, or measurement outcomes) to support noise modelling and analytic sensitivity analysis, and (iii) support for dynamic circuits with classical feedforward, postselection, and
capacity to explore circuit branches at will for debugging.
In practice, many simulators trade one of these requirements for another, e.g., prioritising fast sampling over exact expectation values, restricting noise models to remain tractable, or losing symbolic structure by committing early to numerical parameter values, making it difficult to obtain LERs and circuit-level diagnostics free of Monte Carlo shot noise in the regimes most relevant to QEC.

To address some of the limitations of existing tools, we introduce SyQMA (\textbf{\underline{Sy}}mbolic \textbf{\underline{Q}}uantum \textbf{\underline{M}}emory-efficient \textbf{\underline{A}}nalyser), a classical simulator of universal quantum programs built on an intuitive extension of the stabiliser tableau. For each (generally non-Clifford) arbitrary-angle Pauli rotation and incoherent Pauli-flip channel, SyQMA introduces an auxiliary qubit and new operators with clear properties that allow it to compactly represent the effect of these operations with a tableau. The resulting state representation requires only polynomial memory and time in the circuit size.
The evaluation of expectation values and measurement probabilities requires the computation of the trace of an exponentially large number of stabiliser combinations in the circuit size.
As such, it incurs exponential time in general, but since these combinations can be computed independently, the memory remains polynomial.
We do not envision this version of SyQMA being used for large-scale simulations that involve many measurements and non-Clifford gates, but rather for the detailed analysis of error propagation (stochastic, biased and coherent), fault-tolerance (overall circuit-distance and syndrome-wise) and decoding (maximum-likelihood upper bound, syndrome postselection) in universal QEC gadgets of small to moderate size.

Exploiting these extensions of the stabiliser formalism, SyQMA achieves the following features:
\begin{itemize}[leftmargin=*]
    \item \textbf{Universality with exact Pauli noise:} SyQMA simulates circuits comprising qubit initialisation, Clifford gates, (generally non-Clifford) arbitrary-angle Pauli rotations, projective Pauli measurements, qubit trace-out, and exact incoherent Pauli channels of arbitrary size, not restricted to two-qubit channels as in Stim~\cite{gidney2021stim}. Coherent noise can be modelled via Pauli rotations.
    
    \item \textbf{Symbolic computation:} Expectation values, measurement probabilities, and logical error rates are returned as closed-form symbolic expressions in the circuit parameters: rotation angles, noise rates, and measurement outcomes. Every error rate in every incoherent Pauli channel is treated independently and symbolically, and the resulting expressions reveal the different contributions from the noisy components of the circuit. This makes SyQMA naturally suited to noise modelling and benchmarking, as it allows one to obtain full analytic expressions for noisy expectation values of interest and to deploy more targeted quantum error mitigation techniques.
    
    \item \textbf{Exact strong simulation:} SyQMA returns exact expectation values and measurement probabilities, free from Monte Carlo shot noise. This is particularly valuable for simulating rare events, such as small LERs inaccessible to Monte Carlo methods. Even though there exist modified Monte Carlo methods for more accurate sampling in the low-error regime~\cite{bravyi2013simulation}, the approximations they use produce a bias in the LER that does not exist in our exact simulator.
    
    \item \textbf{Dynamic circuit simulation:}
    SyQMA can simulate dynamic quantum programs in which subsequent circuit operations depend on prior measurement outcomes, such as flagged QEC protocols \cite{chao2018quantum, ryan2021realization, Poor2025ultralow}. It can store measurement outputs symbolically in static circuits and force the simulation into any desired circuit branch (even extremely unlikely ones) created by the intermediate measurements.
    
    \item \textbf{Polynomial memory:} The state representation and all computations require only polynomial memory in the circuit size. Memory becomes exponential only in the number of symbolic parameters retained when extracting information, but for typical analysis in QEC, such as evaluating the LER as a function of one or two symbolic noise rates that parametrise all the noise channels in the circuit, the memory remains polynomial in the circuit size. 
    
    \item \textbf{Circuit-level maximum-likelihood (CL-ML) decoding:} SyQMA computes exact and symbolic CL-ML look-up tables (LUTs) for stabiliser and magic state preparation programs. These CL-ML-LUTs can be computed as analytic functions of the noise parameters, so they need to be computed only once for any noise model within the given parametrisation. 
    
    \item \textbf{Syndrome analysis:} Based on the LUTs, SyQMA can analyse each syndrome individually and determine the LER after applying the CL-ML correction.
    We can also decide on whether the syndrome can be decoded and corrected reliably for various physical noise levels, similar to an exact complementary gap calculation~\cite{gidney2025yoked}.
    This extra information can be used to, e.g., discard upon observing syndromes with an excessive post-correction LER or unreliable correction.
    
    \item \textbf{Exact symbolic logical error rates:} SyQMA computes exact and symbolic expressions for the LER under ideal CL-ML decoding, avoiding the problem of rare-event sampling in Monte Carlo simulations. The symbolic form additionally reveals the most damaging noise sources. 
    
    \item \textbf{Circuit fault-distance verification:} The leading-order asymptotic scaling of the LER in the physical error rate can be read off directly from the symbolic expression, allowing verification that a fault-tolerant state preparation protocol preserves the code distance. To our knowledge, SyQMA is the first tool capable of determining the circuit-level fault distance of a non-Clifford circuit such as magic state preparation.
    
    \item \textbf{Natural to work by hand:} The framework that we introduce provides a natural way to introduce and follow the errors, Pauli rotations and measurement outcomes across the whole execution of a quantum circuit. We can directly inspect the tableau and find information about the final stabilisers, logical operators and decodability properties. This feature is demonstrated with an example in Section~\ref{sec:example}.
\end{itemize}

\noindent Additionally, we highlight the following result:
\begin{itemize}[leftmargin=*]
    \item \textbf{General Pauli channel decomposition:} Borrowing symbolic expressions from the field of quantum simulation \cite{granet2024analog}, we show how to obtain the independent error probabilities from the disjoint error probabilities of an arbitrary $n$-qubit Pauli channel in a QEC simulation context. This enables the exact creation of detector error models in stabiliser simulation. We introduce the result in Section~\ref{ssec:channel_decomposition} and provide further details in Appendix~\ref{app:wh_transform}.
\end{itemize}

\paragraph{Outline.}
Section~\ref{sec:previous_work} discusses previous approaches to the symbolic simulation of universal and dynamic quantum circuits and compares them to SyQMA.
Section~\ref{sec:background} introduces the notation, the stabiliser tableau, and the basic update rules for qubit initialisation and trace-out, Clifford gates, and Pauli projections.
Section~\ref{sec:formalism} presents SyQMA's framework: the factorisation of Pauli noise channels, the update rules for the exact implementation of Pauli-flip channels and Pauli rotations, and the computation of traces, expectation values, and measurement probabilities.
Section~\ref{sec:complexity} analyses the time and memory complexity of the simulator.
Section~\ref{sec:qec} describes the QEC methods: ideal decoding, logical error rate computation, and fault-distance verification.
Section~\ref{sec:example} works through a detailed example on a 3-qubit repetition code.
Section~\ref{sec:results} analyses the fault-tolerant preparation of stabiliser and magic states on the Iceberg $\llbracket k{+}2,k,2\rrbracket$, Steane $\llbracket7,1,3\rrbracket$, Reed-M\"uller $\llbracket15,1,3\rrbracket$, and $\llbracket17,1,5\rrbracket$ codes. 
Section~\ref{sec:outlook} discusses future directions.

For readers interested in the theory of the simulation framework, we recommend sections~\ref{sec:background}-\ref{sec:complexity}, while for readers solely interested in the QEC implications, we recommend sections~\ref{sec:qec}-\ref{sec:results}.

\paragraph{Source code.}
The SyQMA code is publicly available and open-source at \url{https://github.com/Quantinuum/SyQMA}.

\section{Previous work} \label{sec:previous_work}
The landscape of classical simulation of noisy, universal circuits, particularly applied to QEC, has diversified considerably in recent years, and we now briefly survey the main families of methods that are most relevant to our work.

\textit{Stabiliser simulators.}
The stabiliser tableau of Aaronson and Gottesman~\cite{aaronson2004improved} enables the efficient simulation of Clifford circuits, and Gidney's Stim~\cite{gidney2021stim} brought this capability to high performance with fast sampling and detector error model (DEM) extraction.
Some subsequent works have explored alternative frame representations that trade longer initial state construction for faster sampling, either by extracting the relationships between error flips and measurements through backward circuit propagation~\cite{gidney2021stim, delfosse2023simulation}, or by propagating forward symbolic Pauli gates and intermediate measurement results~\cite{fang2024symphase}. Other ideas involve frame representations that explicitly track the noise channels within the tableau~\cite{miller2018propagation, mor2023noisy}. A related line of work exploits the fact that a coherent error can be viewed as a small unitary rotation, and models its effect perturbatively as a Hamiltonian error generator propagated through a Clifford circuit \cite{miller2025efficient, hines2026simulating}.
While these simulators scale to large qubit counts, unlike SyQMA, they are restricted to Clifford operations and cannot handle the non-Clifford gates required for universal computation.

\textit{Extended stabiliser methods.}
To go beyond Clifford circuits, the extended stabiliser formalism decomposes quantum states as sums of stabiliser states~\cite{yoder2012generalization, bravyi2019simulation, calpin2020exploring}, enabling the simulation of circuits with a moderate number of non-Clifford gates.
The most recent contributions on stabiliser-based non-Clifford-augmentation \cite{wan2025cutting, li2025soft, surti2025efficient} have been prompted in particular by the introduction of magic state cultivation \cite{gidney2024magic} as a cheaper alternative to magic state distillation on surface codes.
These methods can handle large qubit counts with few non-Clifford gates and usually rely on ZX-calculus decompositions \cite{kissinger2022classical}, but are typically limited to weak simulation (sampling) and do not provide symbolic or exact expectation values under noise.
Compared to extended stabiliser simulators, SyQMA performs strong simulation of noisy non-Clifford circuits rather than only sampling the noise and the final observable outcomes.

\textit{Tensor network methods.}
Tensor networks \cite{berezutskii2025tensor} offer a flexible framework that can handle both large numbers of qubits and non-Clifford gates, particularly when the entanglement structure of the circuit is favourable. The stabiliser tableau has been interfaced and hybridised with tensor network representations to enhance the amounts of both stabiliserness and magic that can be captured \cite{masot2024stabilizer, nakhl2024stabilizer}. They have also been applied to maximum-likelihood decoding in QEC~\cite{bravyi2014efficient, chubb2021general, bohdanowicz2022quantum, piveteau2024tensor}.
However, tensor network contractions are generally approximate and the computational cost is controlled by the bond dimension, which can grow exponentially with entanglement.
Compared to tensor network simulators, SyQMA can compute expectation values and measurement probabilities with polynomial memory as it is not limited by the depth of the entanglement.

\textit{Pauli propagation.}
A family of Pauli propagation (or backpropagation) methods~\cite{delfosse2023simulation, fontana2023classical, rudolph2023classical, angrisani2024classically, cirstoiu2024fourier, angrisani2025simulating, rudolph2025pauli} computes expectation values by propagating observables in the Heisenberg-picture backwards through the circuit.
Each non-Clifford gate splits a Pauli operator into two or three paths, leading to an exponential sum that can be truncated for approximate simulation.
These methods provide symbolic expressions in the rotation angles and can handle noisy circuits, but they currently cannot store measurement outcomes symbolically or simulate dynamic circuits where subsequent operations depend on prior measurement outputs.
Compared to Pauli path propagation simulators, SyQMA can store measurement outputs symbolically in static circuits and simulate dynamic circuits with classical control. 

\textit{Statevector and density matrix simulators.}
Full statevector and density matrix methods are exact and handle arbitrary gate sets, but their memory requirements scale exponentially in the number of qubits ($2^n$ and $4^n$ respectively), limiting them to small system sizes. 
Compared to density matrix simulators, SyQMA requires only polynomial memory while also treating noise channels exactly and without Monte Carlo shot noise.

\textit{Symbolic simulators.}
A symbolic simulator can produce an expression for a quantity of interest, usually an expectation value, as a function of circuit parameters such as projective measurement results or rotation angles. Obtaining a landscape of expectation values as a function of rotation angles has been particularly useful for the simulation of variational algorithms and the optimisation of their ansatze~\cite{fontana2022spectral, fontana2022efficient}. In the context of QEC, several works focused on the symbolic verification of fault-tolerance properties~\cite{Fang2024, chen2025verifying}. Going beyond the capabilities of these symbolic simulators, SyQMA can obtain symbolic expressions for expectation values as functions of both rotation angles and measurement results, and additionally of the physical error rates of all noise channels.

\textit{Noise decomposition and decoding tools.}
A related aspect for efficient noisy simulations for QEC is the translation of the noisy circuit into a decoding graph (detector error model (DEM)) that adequately captures the correlations of the circuit-level errors. In particular, we present the decomposition of Pauli error channels into independent errors for constructing DEMs, which is essential for both decoding and sampling \cite{algassertDecorrelatedDepolarization,stackexchangeCalculatingWeights,stackexchangeNQubitDepolarizing,stackexchangeStimError,stackexchangeWhereLines,higgott2023improved,derks2025designing, blume2025estimating}.
A general symbolic solution to the problem of converting disjoint probabilities to independent probabilities for a Pauli channel can be found outside the literature of classical simulation of QEC circuits in \cite{granet2024analog}, and in an experimental context for correlated decoding and noise characterisation in \cite{remm2026experimentally, zheng2026efficient}. The symbolic expressions obtained can be linked to the Walsh-Hadamard transform, which relates Pauli error probabilities to Pauli channel eigenvalues. Additional details can be found in Section~\ref{ssec:channel_decomposition}.

SyQMA is thus uniquely positioned to fill a gap in the literature for a simulator that performs exact, noisy, symbolic simulation of dynamic and universal quantum circuits, without being limited by the number of qubits in the system or the Monte Carlo approximation.

\section{Background and notation}
\label{sec:background}
This section introduces the basic notation and definitions used throughout the manuscript. We then present the stabiliser tableau representation of quantum states and a general definition of quantum programs encompassing dynamic circuits, as commonly encountered in QEC. Finally, we summarise the action of Clifford gates, Pauli measurements, and qubit trace-out on the tableau.

\subsection{Notation}
We denote the set of qubits as $\mathcal{Q}$, the trace over a single qubit $q\in\mathcal{Q}$ as $\tr_q$, and the trace over all qubits as $\tr_\mathcal{Q} = \tr_{q_1}\tr_{q_2}\cdots$ for all qubits in $\mathcal{Q}$.
Single-qubit Pauli operators (including the identity) acting on any qubit $q\in\mathcal{Q}$ are $P_q \in \mathcal{P}_q = \{I_q, X_q, Y_q, Z_q\}$.
Multi-qubit Pauli operators $P =\bigotimes_{q\in\mathcal{Q}}P_q$ belong to the set $\mathcal{P}_\mathcal{Q}=\bigotimes_{q\in\mathcal{Q}}\mathcal{P}_q$.
It is convenient to define symbolic Pauli operators as functions of the binary vectors $\bm{x},\bm{z} \in \{\pm1\}^{|\mathcal{Q}|}$:
\begin{equation}
    P(\bm{x},\bm{z}) = \bigotimes_{q\in\mathcal{Q}} X_q^{(1-x_q)/2} Z_q^{(1-z_q)/2} \in \{\pm1, \pm i\}\times \mathcal{P}_\mathcal{Q}.
\end{equation}

For convenience, Pauli operators will be written without the tensor product symbol $\otimes$, including only their non-identity elements, e.g., $X_{q_1}Z_{q_2} \equiv X_{q_1}\otimes Z_{q_2}\otimes I_{q_3}$ for $\mathcal{Q} = \{q_1, q_2, q_3\}$. The tensor product of identity operators is denoted as $I \equiv \bigotimes_{q\in\mathcal{Q}}I_q$. A Pauli operator $P\in\mathcal{P}_\mathcal{Q}$ has trivial support on qubit $q\in\mathcal{Q}$ if it acts as $I_q$ on that qubit. The set of qubits where $P$ acts non-trivially is its support. For a subset $\mathcal{Q}'\subset\mathcal{Q}$, we define $P|_{\mathcal{Q}'} = \bigotimes_{q\in\mathcal{Q}'}P_q$ as the action of $P$ on $\mathcal{Q}'$. The weight of a Pauli operator is the number of qubits in its support.

The \textit{projector} onto the eigenspace $\lambda\in{\pm1}$ of a Pauli operator $P\in\mathcal{P}_\mathcal{Q}$ is $\Pi(\lambda P) = (I+\lambda P)/2$. Similarly, for a vector $\bm{P} = (P_i\in\mathcal{P}_\mathcal{Q})$ of (possibly repeated or anticommuting) $|\bm{P}|$ Pauli operators, the projector onto the subspace defined by the vector $\bm{\lambda} = (\lambda_i \in {\pm1}: \: \forall \, i\in[|\bm{P}|])$ is denoted as the ordered product $\Pi(\bm{\lambda}\bm{P})=\Pi(\lambda_{|\bm{P}|}P_{|\bm{P}|})\cdots\Pi(\lambda_1P_1)$.
We denote some sets as $[k]=\{1,2, \ldots,k\}$, and write element-wise products of two vectors of the same length as $\bm{\lambda}\bm{P} = (\lambda_iP_i:\: \forall\, i\in[|\bm{P}|])$.

\subsection{Stabiliser tableau}
In this work, a \textit{tableau}, defined on a set of qubits $\mathcal{Q}$, is the element-wise product $\bm{T} = \bm{\lambda}\bm{P}$ of a vector $\bm{P}$ of commuting Pauli operators ($[P,P']=0$ for all $P,P'\in\bm{P}$) and a binary vector $\bm{\lambda} \in \{\pm1\}^{|\bm{P}|}$ of the same length.
The tableau represents the (generally unnormalised) state $\rho=w\,\Pi(\bm{T}) = w\,2^{-|\bm{T}|}\sum_{T\in\braket{\bm{T}}} T$, or equivalently, the sum over all commuting elements in the Abelian group $\braket{\bm{T}} = \prod_{T\in\bm{T}}\{I, T\}$ that it generates.
The \textit{normalisation weight} $w$ keeps the norm of the state bounded, $0 \leq \tr_\mathcal{Q}(\rho) \leq 1$, so that it corresponds to the probability of producing the state. 
A state is normalised when $\tr_\mathcal{Q}(\rho)=1$, i.e., when the program produces it deterministically (with unit probability).
The empty tableau $\bm{T}=\emptyset$ represents the completely mixed state $\rho=w\,\Pi(\emptyset)=I/2^{|\mathcal{Q}|}$.

The following transformations preserve the represented state, so we say they produce equivalent tableaus:
\begin{enumerate}
    \item Reordering of elements: $(T_1,\, T_2, \ldots) \equiv (T_2,\, T_1, \ldots)$,
    \item Multiplying any element by another: $(T_1,\, T_2, \ldots) \equiv (T_1T_2,\, T_2, \ldots)$,
    \item Removing linearly dependent elements, i.e., those obtainable as products of others, e.g.,\\
    $(T_1,\, T_2,\, T_1T_2,\ldots) \equiv (T_1,\, T_2,\ldots)$ or trivially $(T_1,\, T_1,\ldots) \equiv (T_1,\ldots)$.
\end{enumerate}
Reducing a tableau to its minimum size can be performed in polynomial time via Gaussian elimination modulo 2 by representing the tableau in symplectic form~\cite{aaronson2004improved}. When reduced, dependent elements become $I$ or $-I$. Elements $I$ can be removed without affecting the represented state. Any element $-I$ produces the \textit{impossible state} $\rho=0$, indicating that the probability of producing such a state is zero, e.g., projecting the $+1$ eigenstate of $Z$ onto its $-1$ eigenstate. Reaching an impossible state terminates the simulation, as it may signal a high-level bug in the program specification.

\subsection{Quantum programs}
The quantum operations considered here are qubit initialisation, Clifford gates, Pauli projection, qubit trace-out, (generally non-Clifford) arbitrary-angle Pauli rotations, and incoherent Pauli-flip channels, possibly controlled by a classical process. These operations are sufficient to describe the evolution of universal and dynamic quantum programs.

A \textit{quantum program} $\mathsf{P}$ is a sequence of classically-controlled quantum operations. The operations may be conditioned on the measurement outputs of earlier Pauli projections in the program.
The \textit{circuit branches} $\mathsf{P}_{\bm{m}}$ are the fixed sequences of circuit operations specified by the program for every pattern of measurement outputs $\bm{m}$. 
Note that quantum programs may apply additional Pauli projections that produce more measurement outputs, so the length of $\bm{m}$ might not be the same for every branch.
Circuit branches that contain no Pauli projections that produce measurement outputs have an empty $\bm{m}=\emptyset$.
Circuit branches map an input quantum state $\rho$ into an output quantum state $\mathsf{P}_{\bm{m}}(\rho)$.
The first program applied on a system with no qubits takes no input, so we denote the output state as $\rho_{\bm{m}}=\mathsf{P}_{\bm{m}}()$ for simplicity.

A quantum program must specify an operation for every possible measurement outcome, otherwise the simulation could not proceed.
One possible action is to halt the computation and possibly restart it when this happens, which would be represented by the tracing-out of all qubits, followed by their reinitialisation and subsequent quantum operations.

\textit{Static programs} are those quantum programs where the operations do not depend on previous measurement outcomes, so circuit branches only differ on the measurement outputs, but not on the operations applied.
Static programs are fixed sequences of quantum operations like those we typically draw in a circuit.
\textit{Dynamic programs} are those where the circuit operations depend on the measurement outcomes, so branches can be very different.
Dynamic programs are very common in QEC and they might look simple, like a sequence of Clifford gates and Pauli measurements that detect noise, followed by a Pauli correction derived by decoding the measurement outputs.
But they can also be more complex, like flagged QEC circuits~\cite{reichardt2020fault, ryan2021realization, Poor2025ultralow} where additional Clifford gates, Pauli measurements and Pauli corrections might be applied depending on the first measurement outputs.

\subsection{Stabiliser state evolution} \label{ssec:basic_evol}
We now proceed to describe how to update the stabiliser tableau to reflect the effect of the qubit initialisation, Clifford gates, projective Pauli measurements, and qubit trace-out. 
These operations alone do not contribute to the exponential simulation cost and are already implemented in stabiliser simulators.
We leave our particular update rule for Pauli rotations and Pauli-flip channels for Section~\ref{sec:formalism}.
All the update rules presented in this work require only polynomial time and memory.

Any quantum program starts by initialising the qubits upon which the rest of the quantum operations will act, and many quantum programs (especially in QEC) initialise qubits during their execution.
\begin{boxenv}{Qubit initialisation} \label{box:qubit_init}
    Initialise a qubit $q \notin \mathcal{Q}$ in the eigenstate $\lambda\in\{\pm1\}$ of a Pauli operator $P_q\in\mathcal{P}_q$.
    It transforms the state $\rho$ defined on $\mathcal{Q}$ into the state $\rho\otimes \Pi(\lambda P_q)$ defined on the qubit set $\mathcal{Q}\sqcup\{q\}$. 
    To obtain the tableau of the transformed state, add a new element $T=\lambda P_q$ to the tableau and the qubit $q$ to $\mathcal{Q}$.
    For $P_q = I_q$, just update the normalisation weight $w$ to $w/2$ to preserve the norm of the state. 
\end{boxenv}

One of the most attractive features of tableaus is that the update rule for Clifford gates is efficient and does not contribute directly to the exponential computation runtime.
We additionally consider the application of symbolic Pauli operators, whose symbolic parameters may depend on previous measurement outputs in the circuit. 
\begin{boxenv}{Clifford gate} \label{box:clifford}
    A Clifford gate $C$ transforms the state $\rho$ into $C\rho C^\dagger$. 
    To obtain the tableau of the transformed state, replace every element $T \in \bm{T}$ by the same element conjugated with the Clifford gate, i.e., by $CTC^\dagger$.
    If the Clifford gate is a Pauli operator $C \in\mathcal{P}_\mathcal{Q}$, every element that anticommutes with it gets multiplied by $-1$.
    If it is a symbolic Pauli operator $C=P(\bm{x}',\bm{z}')$ every element $T = \lambda P(\bm{x},\bm{z}) \in \bm{T}$ gets multiplied by $\prod_{q\in\mathcal{Q}}(-1)^{(1-x_q)(1-z'_q)/4 + (1-z_q)(1-x'_q)/4}$. 
\end{boxenv}

The projection into the eigenspace of a Pauli operator also has a convenient update rule in the stabiliser formalism:
\begin{boxenv}{Pauli projection} \label{box:projection}
    The projection of the state $\rho$, defined on $\mathcal{Q}$, onto the $m\in\{\pm1\}$ eigenspace of a Pauli operator $P \in \mathcal{P}_\mathcal{Q}\setminus\{I\}$ transforms the state into the state $\Pi(m P)\, \rho\, \Pi(m P)$.
    To update the tableau, follow the next steps:
    \begin{enumerate}
    \item Minimise the number of elements in the tableau that anticommute with $P$. If all of them commute, skip this step. To minimise their number:
        \begin{enumerate}
        \item Pick one element that anticommutes with $P$, say $T_1 \in \bm{T}$ to be the \textit{pivot} element.
        \item Multiply every other anticommuting element with it, i.e., replace every other $T_i \in \bm{T}\setminus\{T_1\}$ with $\{P,T_i\}=0$ by $T_1T_i$.
        \item Remove the pivot $T_1$ from the tableau.
        \item Update the normalisation weight $w$ into $w/2$.
        \end{enumerate}
    \item Add a new element $m P$ to the tableau.
    \end{enumerate}
\end{boxenv}
The projection is \textit{deterministic} if the Pauli operator $mP$ or $-mP$ is in the Abelian group $\braket{\bm{T}}$.
In the first case, the projection acts trivially on the state and in the second case, it makes the state impossible.

Finally, tracing-out a qubit has the operational meaning of ``forgetting'' or losing the information on that qubit. 
Tracing-out qubits removes them from the state representation and may remove some degrees of freedom attached to them, which may accelerate the computation of information from the state. 
\begin{boxenv}{Trace-out} \label{box:trace-out}
    Tracing-out a qubit $q\in\mathcal{Q}$ from the state $\rho$, defined on $\mathcal{Q}$, transforms it into the state $\tr_q(\rho)$, defined on $\mathcal{Q}\setminus\{q\}$. To update the tableau, follow the next steps:
    \begin{enumerate}
    \item If all elements in the tableau have trivial support on $q$, update the normalisation weight $w$ into $2w$ and remove $q$ from $\mathcal{Q}$.
    \item Otherwise, minimise the number of elements in the tableau that act non-trivially on $q$. To minimise their number:
        \begin{enumerate}
        \item Pick an element in the tableau that acts non-trivially on $q$, say, $T_1$, to be the \textit{first pivot}. 
        \item Multiply every other element with a non-trivial action on $q$ by the first pivot.
        \item If there are no more elements other than the first pivot that act non-trivially on $q$, proceed to the last step.
        \item Otherwise, select one of them, say $T_2$, to be the \textit{second pivot}.
        \item Multiply every other element that acts non-trivially on $q$ by the suitable combination of the pivots that makes the product act trivially on $q$.
        \item Remove the pivots from the tableau and $q$ from $\mathcal{Q}$. If two pivots are removed, update the normalisation weight $w$ into $w/2$. 
        \end{enumerate}
    \end{enumerate}
\end{boxenv}

\section{SyQMA's framework} 
\label{sec:formalism}
This section formalises our proposed update rule for (generally non-Clifford) arbitrary-angle Pauli rotations and the Pauli-flip quantum operations.
We start by showing how general multi-Pauli noise channels can be decomposed into a composition of Pauli-flip channels, which are amenable to SyQMA's framework.
We then show how the exact (without approximations) application of Pauli rotations and Pauli-flip channels can be made while preserving the tableau framework.
The update rules can be executed with polynomial time and memory due to the introduction of rotation qubits, COS operators, flip qubits, and flip operators. 
Finally, the section describes the computation of information from the state in the SyQMA formalism: expectation values and measurement probabilities.

\subsection{Decomposition of Pauli channels into Pauli-flip channels} \label{ssec:channel_decomposition}
SyQMA represents incoherent noise via individual Pauli-flip channels, each of which flips the state by a single Pauli operator with some probability. 
In this subsection, we show how a general multi-Pauli channel can be exactly decomposed into a composition of such flip channels and provide the closed-form expression for the parameters. We provide a further derivation in Appendix~\ref{app:wh_transform}, showing that this decomposition is naturally governed by the Pauli transfer matrix (PTM) eigenvalues of the channel and the Walsh-Hadamard transform.

A \textit{multi-Pauli channel} $\mathcal{E}$ on a set of $k$ qubits $\mathcal{Q}$ applies every Pauli operator $P \in \mathcal{P}_{\mathcal{Q}}$ (including the identity) with probability $p_P \geq 0$, $\sum_{P} p_P = 1$:
\begin{equation} \label{eq:multi_pauli_channel}
    \mathcal{E}(\rho) = \sum_{P \in \mathcal{P}_{\mathcal{Q}}} p_P \, P \rho P.
\end{equation}
The probabilities $\{p_P\}$ are the \textit{disjoint} error probabilities: each $p_P$ is the probability that \textit{only} the Pauli $P$ is applied. 
In SyQMA, the tableau update rule for incoherent noise relies on a single Pauli-flip channel $\mathcal{E}^{p'_P}_P(\rho) = (1 - p'_P)\,\rho + p'_P\, P\rho P$, so we seek to express $\mathcal{E}$ as a composition of such flip channels:
\begin{equation} \label{eq:channel_decomposition}
    \mathcal{E} = \bigcirc_{P \in \mathcal{P}_{\mathcal{Q}} \setminus \{I\}} \mathcal{E}^{p'_P}_P,
\end{equation}
where the $p'_P$ are \textit{quasi-probabilities} (they may not lie in $[0,1]$), chosen so that the composed channel equals the original one. 

To determine these parameters, we define the commutation indicator $\eta(P,R) = 0$ if $[P,R]=0$ and $\eta(P,R) = 1$ if $\{P,R\}=0$. For any Pauli $P$, let $\mathcal{A}(P) = \{R \in \mathcal{P}_{\mathcal{Q}} : \eta(P,R) = 1\}$ denote its anticommuting set and $\mathcal{C}(P)$ denote its commuting set. The quasi-probabilities for the individual flip channels can be written directly in terms of the initial disjoint probabilities as:
\begin{equation} \label{eq:flip_factor_from_probs}
    p'_P = \frac{1}{2} - \frac{1}{2} \left( \frac{\displaystyle\prod_{R \in \mathcal{A}(P)} \left[ 1 - 2\sum_{P' \in \mathcal{A}(R)} p_{P'} \right]}{\displaystyle\prod_{R \in \mathcal{C}(P)} \left[ 1 - 2\sum_{P' \in \mathcal{A}(R)} p_{P'} \right]} \right)^{2/4^{k}}.
\end{equation}
For example, the single-qubit depolarising channel $\mathcal{E}(\rho) = (1-p')\rho + p'_X X\rho X + p'_Y Y\rho Y + p'_Z Z\rho Z$ is decomposed into $X$-, $Y$-, and $Z$-flip channels with flip parameters obtained from Eq.~\eqref{eq:flip_factor_from_probs}. 

The quasi-probabilities $p'_P$ may lie outside $[0,1]$, and in fact it is stated in \cite{zheng2026efficient} that they can have values in $[-\infty, 1/2]$. Negative quasi-probabilities can happen particularly in the case of biased noise. For example, if a single-qubit channel has $X$ and $Z$ errors, but no $Y$ errors, then the factorisation would result in a $Y$-channel with negative quasi-probability to account for that effect.
However, even if we obtain unphysical independent error channels, their composition as the original channel is still physical, so this does not create a problem for calculations with SyQMA. For physical noise at small error rates, $p'_P$ are valid probabilities. 

This closed-form relation was derived for the first time in \cite{granet2024analog}, and subsequently in \cite{remm2026experimentally, zheng2026efficient}, and Eq.~\eqref{eq:flip_factor_from_probs} provides a general symbolic expression for the \textit{independent} (flip-channel) error probabilities from the \textit{disjoint} probabilities of an arbitrary $n$-qubit Pauli channel.
Previous works had noted the difficulty of this conversion in general~\cite{higgott2023improved}, with closed-form expressions known only for the depolarising channel~\cite{algassertDecorrelatedDepolarization}.
This decomposition has direct implications for the exact construction of detector error models in Clifford simulation.

\subsection{Tableau update for Pauli-flip channels and Pauli rotations} \label{ssec:new_evolution}
SyQMA's framework introduces auxiliary qubits, a new trace and new operators in order to implement Pauli-flip channels and Pauli rotations while preserving the structure and properties of the tableau.
As explained in Appendix~\ref{app:definitions}, these definitions arise naturally when every Pauli-flip channel and Pauli rotation in a quantum program are injected from auxiliary qubits that are projected onto a mixed state or non-stabiliser pure state, respectively.
Upon absorbing these projections into the definition of a new trace, what remains is an entirely Clifford program that admits a highly natural representation with a tableau framework.
Under the new trace the Pauli operators acting on the auxiliary qubits are no longer traceless, but rather have a trace value related to the flip factor of the Pauli-flip channel or the angle of the Pauli rotation. 
In order to avoid confusion, we introduce new operators as the Pauli operators acting on the auxiliary qubits. 

We first introduce the \textit{set of flip qubits} $\mathcal{F}$ and the \textit{set of rotation qubits} $\mathcal{R}$, which contain those auxiliary qubits that inject Pauli-flip channels or Pauli rotations, respectively.
Together with the computational qubits they form the set of all qubits $\mathcal{N}=\mathcal{Q}\sqcup\mathcal{R}\sqcup\mathcal{F}$. 
For every Pauli-flip channel $\mathcal{E}_P^p$ we add an auxiliary qubit $f$ to $\mathcal{F}$ and write it as a subindex of the error rate $p_f$ and flip factor $\epsilon_f = 1-2p_f$ of that channel.
Similarly, for every Pauli rotation $R_P^\theta$ we add an auxiliary qubit $r$ to $\mathcal{R}$ and write it as a subindex of the angle $\theta_r$ of that rotation.

We then introduce \textit{flip operators} $F$ acting on each flip qubit $f\in\mathcal{F}$ and \textit{COS operators} $C$, $O$, $S$ acting on each rotation qubit $r \in \mathcal{R}$. 
Similar to Pauli operators, they satisfy that $F^2 = C^2 = O^2 = S^2 = I$ and $CO=iS$.
The trace $\tr()$ is replaced by a new trace $\Tr()$ that depends on the qubits it acts upon and provides the flip and COS operators with their key properties: 
\begin{itemize}
    \item \textbf{Computational qubit $q\in\mathcal{Q}$:} $\Tr_q(I_q)=2$ and $\Tr_q(X_q)=\Tr_q(Y_q)=\Tr_q(Z_q)=0$ as usual,
    \item \textbf{Flip qubit $f\in\mathcal{F}$:} $\Tr_f(I_f) = 2$ and $\Tr_f(F_f) = 2\epsilon_f$,
    \item \textbf{Rotation qubit $r\in\mathcal{R}$:} $\Tr_r(I_r) = 2$, $\Tr_r(C_r) = 2\cos(\theta_r)$, $\Tr_r(O_r) = 0$, $\Tr_r(S_r) = 2\sin(\theta_r)$.
\end{itemize}

The following update rules are such that only identity and flip operators act on the flip qubits and only identity and COS operators act on the rotation qubits, so the properties above are all that is needed to update the tableau and recover the correct information when it is evaluated. 
\begin{boxenv}{Pauli-flip channel} \label{box:pauli-flip}
A Pauli-flip channel $\mathcal{E}_P^p$ transforms a state $\rho$ into $\mathcal{E}_P^p(\rho) = (1-p)\rho + p P\rho P$ for some $P\in \mathcal{P}_\mathcal{Q}$ and probability (or quasi-probability) $p$.
To update the tableau, follow the next steps:
\begin{enumerate}
    \item If all elements in the tableau commute with the generator $P$, the state and the tableau are preserved, so skip the next steps. Otherwise:
    \item Add a flip qubit $f$ to the set of flip qubits $\mathcal{F}$.
    \item Multiply all elements in the tableau that anticommute with $P$ by an $F_f$ operator.
    \item Update the normalisation weight $w$ into $w/2$.
\end{enumerate}
\end{boxenv}

And for Pauli rotations the update rule is:
\begin{boxenv}{Pauli rotation} \label{box:rotation}
A (generally non-Clifford) arbitrary-angle Pauli rotation $R^\theta_P = \exp(-i\theta P /2)$ transforms the state $\rho$ into the conjugated state $R^\theta_P \rho R^{-\theta}_P$.
To update the tableau, follow the next steps:
\begin{enumerate}
    \item If all elements in the tableau commute with the generator $P$ of the rotation, the state and tableau are preserved, so skip the next steps. Otherwise:
    \item Add a rotation qubit $r$ to the set of rotation qubits $\mathcal{R}$.
    \item Multiply all elements in the tableau that anticommute with $P$ by a $C_r$ operator.
    \item Add a new element $P\otimes O_r$ to the tableau.
\end{enumerate}
\end{boxenv}
Note that all stabiliser operators in the tableau still commute due to the anticommutation of $C$ and $O$ operators.

\subsection{Canonical form and trace}
Until now we have presented how to describe the evolution of a state under various quantum operations as modifications to the tableau that represents it. 
But to compute information such as expectation values, measurement probabilities, overlaps between two states or measurement probabilities on the measurements, one needs to compute the trace of the state $\tr(P\rho)$ up to a Pauli operator $P\in \mathcal{P}_\mathcal{Q}$.

Before computing the trace, it is convenient to transform the tableau into its \textit{canonical form} $\bm{T} = (\bm{T}_\mathcal{Q},\, \bm{T}_{\mathcal{R}},\,  \bm{T}_{\mathcal{F}},\, \bm{T}_\emptyset)$.
This is obtained by writing the tableau in its symplectic form, with qubits in columns ordered by their type $\mathcal{Q}$, $\mathcal{R}$ and $\mathcal{F}$, and performing Gaussian elimination on the rows.
The canonical form is then such that $\bm{T}_\mathcal{Q},\, \bm{T}_{\mathcal{R}},\,  \bm{T}_{\mathcal{F}}$ contain only elements that cannot be obtained as the product of others and
\begin{itemize}
    \item $\bm{T}_\mathcal{Q}$ elements have non-trivial support on $\mathcal{Q}$,
    \item $\bm{T}_{\mathcal{R}}$ elements have trivial support on $\mathcal{Q}$ but non-trivial support on $\mathcal{R}$, 
    \item $\bm{T}_{\mathcal{F}}$ elements have trivial support on $\mathcal{Q}\sqcup\mathcal{R}$ but non-trivial support on $\mathcal{F}$, 
    \item $\bm{T}_\emptyset$ contains the rest of the elements, which have trivial support on the entire qubit set $\mathcal{N}$ and are of the form $b I$, i.e., the product of binary functions $b\equiv b(\bm{\lambda}, \bm{m})\in\{\pm1\}$ of the eigenvalues and measurement outputs and the global identity.
\end{itemize}
The canonical form makes evident the set of constraints $\mathcal{C}=\left\{b\neq -1: \: \forall \, b I \in \bm{T}_\emptyset\right\}$. If any constraint is not satisfied, the state is impossible (unphysical).
The canonical form also helps speeding-up the computation of the trace by moving as many elements as possible to $\bm{T}_\emptyset$.

Once the canonical form is obtained, computing the trace is faster and clearer.
\begin{boxenv}{Trace} \label{box:trace}
To compute the trace $\Tr(P\rho)$ of the state $\rho$ defined on $\mathcal{N}=\mathcal{Q}\sqcup\mathcal{R}\sqcup\mathcal{F}$, multiplied by the Pauli operator $P\in\mathcal{P}_\mathcal{Q}$, follow the next steps:
\begin{enumerate}
    \item Get the tableau in the canonical form: $\bm{T} = (\bm{T}_{\mathcal{Q}},\, \bm{T}_{\mathcal{R}},\, \bm{T}_{\mathcal{F}},\, \bm{T}_{\emptyset})$, with the constraints $\mathcal{C}$.
    \item If any constraint is not satisfied, the state is impossible and the trace is trivially zero. Otherwise continue.
    \item Find a combination $\bm{T}'_\mathcal{Q} \subset \bm{T}_\mathcal{Q}$ such that the Pauli part of their product is $\prod_{T'\in\bm{T}'_\mathcal{Q}} T'|_\mathcal{Q} = P$. If such a combination exists, it is unique thanks to the canonical form. Otherwise, such a combination does not exist and the trace is trivially zero. Whether it exists or not, it can be found in polynomial time via Gaussian elimination of the elements in $\bm{T}_\mathcal{Q}$.
    \item Compute such a product as $T_{\mathcal{Q}}= \prod_{T'\in\bm{T}'_\mathcal{Q}}T'$.
    \item The trace is then a sum of traces over products of stabiliser operators:
    \begin{equation} \label{eq:trace}
        \Tr(P\rho) = \frac{w}{2^{|\bm{T}_\mathcal{R}|+|\bm{T}_\mathcal{F}|}}\sum_{T\in\braket{\bm{T}_{\mathcal{R}}, \bm{T}_{\mathcal{F}}}}\Tr(T_{\mathcal{Q}}\,T)\prod_{T=b I\in\bm{T}_\emptyset}(1+b)/2.
    \end{equation}
\end{enumerate}
\end{boxenv}
Therefore, in the worst-case scenario, computing the trace requires expanding the group $\braket{\bm{T}_{\mathcal{R}}, \bm{T}_{\mathcal{F}}}$, which is of size $2^{|\bm{T}_{\mathcal{R}}| + |\bm{T}_{\mathcal{F}}|}$. 
Since the terms in the trace can be computed independently from each other, the computation of the trace can be executed with polynomial memory and large amounts of parallelisation.
Regarding the computation time, we discuss in Section~\ref{sec:complexity} that the size of $|\bm{T}_{\mathcal{R}}|$ is much smaller than the number $|\mathcal{R}|$ of Pauli rotations and that $|\bm{T}_{\mathcal{F}}|$ is upper-bounded by the number of projections that are deterministic in the absence of noise rather than on the number $|\mathcal{F}|$ of Pauli-flip channels. 
Additionally many terms in this expansion trivially vanish due to the presence of $O_r$ operators (traceless), so their early identification brings huge computational savings. 
Overall, the computation time is in practice much smaller than the worst-case scenario.

\subsection{Computing expectation values and measurement probabilities}
Equipped with the computation of the trace, we can now compute information from the state.
The \textit{expectation value} of a Pauli operator $P\in \mathcal{P}_\mathcal{Q}$ is $\mathbb{E}[P|\rho] = \Tr(P\rho)/\Tr(\rho)$.
If any of the constraints $\mathcal{C}$ from the canonical form is not satisfied, the state is $\rho=0$, and there is no need to compute any trace because we know they are all zero.

The \textit{measurement probability} $\mathbb{P}[\bm{P}=\bm{m}|\rho]=\Tr\left(\Pi(\bm{m P})\rho\right)/\Tr(\rho)$ is the probability of sampling a measurement output $\bm{m}\in\{\pm1\}^{|\bm{P}|}$ when a vector $\bm{P}$ of (possibly anticommuting and repeated) Pauli operators is measured on $\rho$.
A common situation is when a subset of qubits $\mathcal{Q}' \subset\mathcal{Q}$ is measured in a \textit{Pauli basis} $\bm{P} =(P_q \in \{X_q, Y_q, Z_q\}: \: \forall \, q\in\mathcal{Q}')$.
If any constraint is unsatisfied, then $\rho=0$, so all measurement probabilities are zero.

We can also compute the more general \textit{transition probability} $\mathbb{P}[\mathsf{P}_{\bm{m}}(\rho)|\rho] = \Tr(\mathsf{P}_{\bm{m}}(\rho))/\Tr(\rho)$ from an input state $\rho$ to the output state $\mathsf{P}_{\bm{m}}(\rho)$ of a circuit branch $\mathsf{P}_{\bm{m}}$ with measurement outputs $\bm{m}$. 
Importantly, each trace is taken over the qubit set on which each state is defined, and if any constraint from any of the two states is not satisfied, the transition probability is zero.
The transition probability coincides with the measurement probability, $\mathbb{P}[\mathsf{P}_{\bm{m}}(\rho)|\rho]=\mathbb{P}[\bm{P}=\bm{m}|\rho]$, for any static quantum program $\mathsf{P}$ whose branches $\mathsf{P}_{\bm{m}}$ only apply the measurement sequence $\bm{P}$ to the input state, i.e., they do not contain other quantum operations.
On state preparation programs, that output a (generally unnormalised) state $\rho_{\bm{m}}=\mathsf{P}_{\bm{m}}()$ without input state $\mathbb{P}[\rho_{\bm{m}}] = \Tr(\rho_{\bm{m}})$ is the probability of the measurement output $\bm{m}$ when executing the state preparation program.

For static quantum programs, the measurement probability or the transition probability can be computed exactly in terms of symbolic measurement outputs $\bm{m}$, but storing this distribution consumes, in the worst-case, memory that is exponential in the number $|\bm{m}|$ of measurement outputs, as we discuss in the next section.
Having access to the entire distribution allows sampling from it using standard statistical techniques~\cite{reuven2008sim_monte}.
Polynomial memory in sampling can be achieved by sampling from marginals because these can be computed without symbolic measurement outputs. 
Instead, one evolves the state until right before the first Pauli measurement $P_1 \in\bm{P}$, obtaining the state $\rho_1$, computes the marginal probability $\mathbb{P}[P_1=+1|\rho_1]$, and samples the first bit $m_1 \in \{\pm1\}$ from it.
The state is then projected onto the eigenstate $m_1$ of $P_1$ and the evolution proceeds until right before the second Pauli measurement $P_2 \in\bm{P}$, where the state $\rho_2$ is obtained, the marginal probability $\mathbb{P}[P_2=+1|\rho_2]$ is computed, and the second bit $m_2 \in\{\pm1\}$ is sampled from it.
Repeating this process $|\bm{m}|$ times returns a complete sample $\bm{m}=(m_1,\, m_2, \ldots, m_{|\bm{m}|})$ from the distribution $\mathbb{P}[\bm{P}=\bm{m}|\rho]$ without the need of storing any symbolic measurement output or consuming exponential memory.

In dynamic circuits, each measurement output can divert the quantum circuit into very different branches, so it is not possible to compute the symbolic distribution $\mathbb{P}[\bm{P}=\bm{m}|\rho]$, given that the evolution of the state depends on the measurement outputs sampled along the way.
Fortunately, sampling from dynamic circuits is perfectly possible and still consumes polynomial memory when using marginals.
The runtime in all these sampling methods is exponential in the circuit size due to the need of computing traces.

\section{Complexity}
\label{sec:complexity}
SyQMA evolves the state in polynomial time and memory in the circuit size, and computes information with polynomial memory, but exponential runtime.
In this section, we provide more detail on the worst-case and typical runtimes in several scenarios. 

The state is evolved by updating the tableau via the update rules in Secs.~\ref{ssec:basic_evol} and~\ref{ssec:new_evolution}, which are analogous to those in stabiliser simulators.
Since stabiliser simulators update the state with a worst-case linear runtime and quadratic memory in the number of qubits~\cite{aaronson2004improved}, we should expect a quadratic worst-case complexity $\mathcal{O}\left(|\mathcal{N}|^2\right)$ in the total number of qubits, i.e., the sum of normal, rotation, and flip qubits.
The runtime to obtain the canonical form is $\mathcal{O}\left(|\mathcal{N}|^3\right)$, since it performs Gaussian elimination, but the real bottleneck is the exponential runtime for computing information from the state.

Before commenting on this exponential runtime, let us highlight one of the main features of SyQMA: computing information requires only polynomial memory in the circuit size.
This can be achieved by computing the traces in Eq.~\eqref{eq:trace} one by one without ever keeping the expanded exponentially-sized set $\braket{\bm{T}_{\mathcal{R}} , \bm{T}_{\mathcal{F}}}$ in memory.
However, the memory is exponential in the number of circuit parameters we keep symbolic.
This is because in the worst-case scenario, the trace of every term from $\braket{\bm{T}_{\mathcal{R}}, \bm{T}_{\mathcal{F}}}$ will be a different product of symbolic variables that cannot, in general, be grouped together. 
Nevertheless, if many circuit parameters depend on a few symbolic variables and we perform a Taylor expansion in these variables, we can start grouping terms and control the memory consumption at the cost of a Taylor approximation error.
For example, in typical QEC programs, one expresses all Pauli noise channels as simple linear functions of one or two symbolic parameters, e.g., the incoherent Pauli errors and the qubit loss rates, or the gate and the memory error rates. 
Importantly, in dynamic circuits, the state evolution requires sampling the measurement outputs that control the set of subsequent quantum operations, so no circuit parameter that affects the measurement probabilities can be kept symbolic.

The runtime for computing the trace scales exponentially as $\mathcal{O}\left(2^{|\bm{T}_{\mathcal{R}}| + |\bm{T}_{\mathcal{F}}|}\right)$. 
However, since many elements of the group $\braket{\bm{T}_{\mathcal{R}}, \bm{T}_{\mathcal{F}}}$ may contain single $O_r$ operators, whose trace is zero, avoiding the computation of their traces can significantly speed up the trace computation.
Additionally, the number of elements in $\bm{T}_{\mathcal{R}}$ is strictly (and even significantly) smaller than the number $|\mathcal{R}|$ of Pauli rotations.
This is easier to understand if we set the tableau in the canonical form before and after every rotation.
Every Pauli rotation with a generator $P\in\mathcal{P}_\mathcal{Q}$ that cannot be obtained as a product of elements in $\{T|_\mathcal{Q}:\: \forall\,T\in\bm{T}_\mathcal{Q}\}$ adds a new element $P\otimes O_r$ to the tableau, but taking the canonical form places it in $\bm{T}_\mathcal{Q}$ rather than in $\bm{T}_\mathcal{R}\sqcup\bm{T}_\mathcal{F}$, so it does not increase the runtime exponentially.
Otherwise, if the generator can be expressed as such product, the new element will be added to $\bm{T}_\mathcal{R}$ after taking the canonical form, where it will contribute to the exponential runtime.

This is particularly powerful in circuits like Trotterised Ising evolution or QAOA~\cite{farhi2014qaoa}, whose quantum programs initialise all qubits in the $\ket{+}$ state and then implement alternating layers of Pauli rotations over two-qubit $ZZ$ gates and then over single-qubit $X$ gates. 
After the qubit initialisation the tableau in canonical form contains only single-qubit $X$ (those that stabilise the initial states $\ket{+}$) elements inside $\bm{T}_\mathcal{Q}$, and one by one, every $ZZ$ generator that cannot be expressed as a product of the ones already in the tableau gets added to $\bm{T}_\mathcal{Q}$ as $ZZ\otimes O_r$ elements, thus not increasing exponentially the computation time of the trace.
Those that can be expressed as a product of the ones already in the tableau do contribute to the exponential runtime.

Moving our attention to $\bm{T}_{\mathcal{F}}$, its size is upper-bounded by the number of Pauli projections that would be deterministic in the absence of Pauli-flip channels, but that due to their presence they become non-deterministic.
Again, consider the tableau is brought into the canonical form before and after the Pauli-flip channel.
Since (unlike Pauli rotations) Pauli-flip channels do not add new elements to the tableau, as they rather append flip operators $F$ to existing elements, these channels cannot add elements to $\bm{T}_{\mathcal{F}}$.
However, a deterministic projection of a Pauli operator $P\in\mathcal{P}_\mathcal{Q}$, i.e., that can be expressed as a product of elements in $\prod_{T\in\bm{T}_\mathcal{Q}} T|_\mathcal{Q}$, adds an element to $\bm{T}_\mathcal{R}$ if the product $\prod_{T\in\bm{T}_\mathcal{Q}} T$ has non-trivial support on $\mathcal{R}$ and to $\bm{T}_\mathcal{F}$ if it has trivial support on $\mathcal{R}$ but non-trivial on $\mathcal{F}$ (and to $\bm{T}_\emptyset$ if it has trivial support on $\mathcal{R}\sqcup\mathcal{F}$).
Otherwise, a non-deterministic projection where $P$ cannot be expressed as such a product adds a new element $mP$ to $\bm{T}_\mathcal{Q}$, where it does not contribute to the exponential runtime. 

No operations other than Pauli rotations and deterministic Pauli projections can add elements to $\bm{T}_\mathcal{R}\sqcup\bm{T}_\mathcal{F}$.
Clifford gates (including Pauli operators) preserve the linear independence of the elements in $\bm{T}_\mathcal{Q}$, so that taking the canonical form cannot move any element from it to $\bm{T}_\mathcal{R}\sqcup\bm{T}_\mathcal{F}$. 
Qubit trace-out only removes elements from the tableau. 
Simple recombinations of the elements in the tableau followed by taking again the canonical form may modify these sets, but they will preserve their size.

\section{QEC methods}
\label{sec:qec}
As advertised in the introduction, SyQMA's capabilities are very convenient for QEC programs.
The section starts with a brief notation-setting on QEC, then describes the generation of the circuit-level maximum-likelihood (CL-ML) decoding look-up table (LUT) for a state preparation program.
As a simple extension we explain how to compute the exact logical error rate (LER) and verify the fault distance of a state preparation program.

\subsection{QEC preliminaries}
Consider a QEC code $\llbracket n,k,d\rrbracket$ that encodes $k$ logical qubits into a set $\mathcal{Q}$ of $n$ physical qubits at distance $d$, i.e., that can detect any multi-qubit Pauli error of integer weight $w \leq d-1$ and correct any error of weight up to $(d-1)/2$.
We consider only odd-distance codes in this work, with the exception of the Iceberg code, with $d=2$.
The \textit{set of stabiliser generators} $\bm{S} \subset\mathcal{P}_\mathcal{Q}$ is made of the minimum number of commuting stabiliser operators that generate the stabiliser group $\braket{\bm{S}}$.
All stabilisers and logical operators commute except for the pairs of logical operators $\Bar{X}_i, \Bar{Z}_i \in \mathcal{P}_\mathcal{Q}$, that anticommute $\{\Bar{X}_i, \Bar{Z}_i\}=0$ for every logical qubit $i=[k]$.
The \textit{code space} $\Pi(\bm{S})$ is the common $+1$ eigenspace of all stabiliser operators.
\textit{Detectable errors} are all quantum operations that take the state out of the code space, while \textit{logical errors} preserve the code space but act non-trivially on it.  
Therefore, detectable errors can be detected by the measurement of the stabiliser generators, whose output $\bm{s}$ is the \textit{syndrome}.

For the purpose of decoding, it is convenient to define the \textit{destabilisers} of the code as the $|\bm{S}|$ multi-qubit Pauli operators $\bm{D} \in \mathcal{P}_\mathcal{Q}$ that anticommute one-to-one with the stabilisers $\{D_i, S_i\}=0$ for all $i=[|\bm{S}|]$, commute among themselves, commute with the rest of the stabilisers, and commute with the logical operators. 
After a noiseless measurement of the syndrome, the state can be returned to the code space by applying the destabilisers corresponding to the stabilisers measured as $-1$.
This preserves the expectation value of the logical operators given their commutation with the destabilisers.

The \textit{logical stabilisers} $\Bar{L}_i$ of a logical state are the $k$ linear combinations of logical operators that act trivially on the logical state.
For example, $\Bar{L}=\Bar{Z}$ for a logical $\ket{\Bar{0}}$, or $\Bar{L}_1=\Bar{Z}_1\Bar{Z}_2$ and $\Bar{L}_2=\Bar{X}_1\Bar{X}_2$ for a logical Bell pair $(\ket{\Bar{0}\Bar{0}}+\ket{\Bar{1}\Bar{1}})/\sqrt{2}$.
For the magic state $\ket{\Bar{H}} = \cos(\pi/8)\ket{\Bar{0}} + \sin(\pi/8)\ket{\Bar{1}}$, the logical stabiliser is $\Bar{L} = (\Bar{X}+\Bar{Z})/\sqrt{2}$. 
Note that since a logical magic state can be produced by acting on a logical stabiliser state via some logical non-Clifford unitary gate $\Bar{G}$, the corresponding logical stabiliser can be brought back to a logical Pauli stabiliser via $\Bar{G}^\dagger$. 
For example, since $\ket{\Bar{H}} = \Bar{G}\ket{\Bar{0}}$ for $\Bar{G} = R^{\pi/4}_{\Bar{Y}}$, then $\Bar{G}^\dagger \left((\Bar{X}+\Bar{Z})/\sqrt{2}\right) \Bar{G} = \Bar{Z}$.
The \textit{logical destabilisers} $\Bar{D}_i$ are logical operators that anticommute with only their corresponding logical stabiliser $\Bar{L}_i$ for every $i=[k]$, commute among themselves, and commute with every other logical stabiliser, stabiliser operator, and destabiliser.
In the examples above, $\Bar{D}=\Bar{X}$ is the logical destabiliser for the logical state $\ket{\Bar{0}}$, $\Bar{D}_1 = \Bar{X}_1$ and $\Bar{D}_2 = \Bar{Z}_2$ are the logical destabilisers for the logical Bell pair, and $\Bar{D} = \Bar{G} \Bar{X} \Bar{G}^\dagger = (\Bar{X}-\Bar{Z})/\sqrt{2}$ is the logical destabiliser for the magic state.

\subsection{Ideal decoding in the simulator}
The task of a state preparation program $\mathsf{P}$ in QEC is to prepare a specific state $\rho$.
For example, noiseless deterministic state preparations, such as the typical surface code initialisation, obtain random measurement outputs $\bm{m}$ (even in the absence of noise) from measuring the stabilisers on an initial product state of $\ket{0}$s where they do not act trivially, and then apply a Pauli correction depending on $\bm{m}$ to deterministically prepare $\rho$.
Analogously, repeat until success state preparation programs alternately implement a state preparation circuit $\mathsf{P}_{\bm{m}_i}$ and a circuit restart (trace-out of all qubits and run the circuit again) until the measurement output is trivial (a string of $+1$'s $\bm{m}_i=\bm{+1}$), at which point the target state $\rho$ is produced, again irrespective of the history of measurement outputs $\bm{m}_i$.

However, a noisy quantum program produces generally different noisy states $\rho_{\bm{m}}=\mathsf{P}_{\bm{m}}()$ for every branch.
Noisy programs can be obtained from a noiseless program by inserting noise channels (small-angle Pauli rotations for coherent noise and Pauli-flip channels for incoherent noise) that, typically, emulate the hardware noise.
Deterministic programs use the measurement outputs $\bm{m}$ to propose a Pauli correction, while non-deterministic programs restart the circuit until the trivial measurement output is obtained.

The goal of a decoder is to maximise the \textit{logical success rate} $\mathrm{LSR}[\rho] = \mathbb{P}[\bm{L}=\bm{+1}|\rho]$ of a state $\rho$, i.e., the probability of measuring in $+1$ all the logical stabilisers.
The \textit{logical error rate} $\mathrm{LER}[\rho] = 1 - \mathrm{LSR}[\rho]$ is the probability of measuring at least one logical stabiliser in $-1$.
More generally, we can define the LSR of a state preparation program by averaging over the branches:
\begin{equation} \label{eq:average_lsr}
    \mathrm{LSR}[\mathsf{P}] = \sum_{\bm{m}}\mathrm{LSR}[\rho_{\bm{m}}] \, \mathbb{P}[\rho_{\bm{m}}].
\end{equation}

SyQMA decodes QEC noisy state preparation programs by simulating an ideal CL-ML decoder, which can be described by the noiseless decoding program and an optimal correction program.
The \textit{decoding program} $\mathsf{D}$ noiselessly projects the input state onto the syndrome $\bm{s} \in \{\pm1\}^{|\bm{S}|}$ of the stabilisers $\bm{S}$, and then applies the product $D(\bm{s}) = \prod_{i=1}^{|\bm{s}|}D_i^{(1-s_i)/2} \in \braket{\bm{D}}$ of the destabilisers $D_i \in\bm{D}$ for the syndrome $\bm{s}$, resulting in the noisy projected and destabilised states $\mathsf{D}_{\bm{s}}(\rho_{\bm{m}}) = D(\bm{s}) \, \Pi(\bm{s S})\,\rho_{\bm{m}} \, \Pi(\bm{s S}) \, D(\bm{s})$.
This state is fully in the code space, i.e., all stabilisers act trivially on it, but it potentially carries a logical noise channel that we must correct.

The \textit{optimal correction program} $\mathsf{C}^*$ applies the optimal circuit branch $\mathsf{C}^*_{\bm{m},\bm{s}}$ for every measurement output $\bm{m}$ and syndrome $\bm{s}$ from a set $\bm{\mathsf{C}}$ of available corrections.
The optimal circuit branches are the ones from the set that maximise LSR on the decoded state:
\begin{equation}
    \mathsf{C}_{\bm{m}, \bm{s}}^* = \operatorname*{arg\,max}_{\mathsf{C} \in \bm{\mathsf{C}}} \mathrm{LSR}[\mathsf{D}_{\bm{s}}(\rho_{\bm{m}})].
\end{equation}
Typically, the available correction quantum programs are those that can be executed without introducing too much additional noise, such as Pauli operators because they can be implemented as a tensor product of single-qubit physical Pauli operators, that introduce little noise, or even implemented in software with zero noise.
But many QEC codes possess other transversal logical gates (like colour codes~\cite{bombin2006colorcode}, possessing a transversal Hadamard) that could be potentially included in the set of correction programs.
The corrected LSR of a state preparation program is therefore $\mathrm{LSR}[\mathsf{C}^*\circ\mathsf{D}\circ\mathsf{P}]$.

\subsection{Verifying the fault distance of a quantum program}
We can also use this decoding framework to verify the fault distance of the state preparation program.
One of the two Gottesman's criteria~\cite{Gott2024surviving} for a quantum state preparation program to be fault-tolerant (FT) is that the distance is preserved.
A state preparation program preserves the distance $d$ of the code if all combinations of up to $t=(d-1)/2$ noise events can be successfully decoded by an ideal decoder.
Therefore, logical errors can be caused only by $t+1$ or more noise events, so in a noise model where the parameters of every noise channel are some $\mathcal{O}(p)$ function of a single parameter $p$, the LER must be $\mathcal{O}\left(p^{t+1}\right)$.
Since SyQMA simulates the state and the decoding with symbolic parameters, the LSR and LER obtained by SyQMA will be symbolic functions of those parameters, so we can then verify this FT criterion by computing the leading order of the Taylor expansion in $p$ of the LER.

Full fault-tolerance verification under Gottesman's criteria requires additional checking that no combination of up to $f \leq t$ noise events of probability $\mathcal{O}(p)$ leads to an error of minimum weight (up to stabiliser operators and logical stabiliser operators) larger than $f$. 
Pauli propagation tools and quick programs for finding the minimum weight of an error should be used for this purpose. 

\subsection{Decoding on hardware}
Now we use the ideal decoder produced in the previous section in a realistic scenario when running on hardware.
If a noisy QEC cycle was to be performed right after the noisy state preparation, this decoder could be used to correct the resulting state.
The approximation made is to assume that the noisy QEC program executed on hardware can extract the noiseless syndrome without introducing additional noise. 
To be more precise, the setup for decoding is the following:
\begin{itemize}
    \item We aim to run a noiseless quantum program on a quantum computer to prepare a unique noiseless state $\rho$, that is the same for all program branches.
    
    \item However, due to noise, the quantum computer runs instead a noisy program that we approximate in our simulator as $\mathsf{P}$ by inserting noise that emulates the hardware noise in the noiseless program.
    
    \item The measurement outputs $\bm{m}$ obtained in the quantum computer determine a program branch $\mathsf{P}_{\bm{m}}$ and lead to the noisy state $\rho_{\bm{m}}=\mathsf{P}_{\bm{m}}()$ in our simulator.
    
    \item To obtain the ideal decoder, we simulate the decoding program $\mathsf{D}$, which, for each branch $\mathsf{D}_{\bm{s}}$ implements a noiseless projection $\Pi(\bm{s S})$ of the stabilisers $\bm{S}$ into a symbolic syndrome $\bm{s}$ and applies the product of destabilisers $D(\bm{s})$. The resulting state $\mathsf{D}_{\bm{s}}(\rho_{\bm{m}})$ is fully inside the code space, but might carry a logical error channel.
    
    \item We find from a set $\bm{\mathsf{C}}$ of available correction programs the optimal correction programs for every branch and syndrome $\mathsf{C}_{\bm{m}, \bm{s}}^*$, i.e, that maximise the probability that all logical stabilisers $\bm{\Bar{L}}$ are measured in $+1$ on the recovered state $\mathsf{C}_{\bm{m}, \bm{s}}^*\circ\mathsf{D}_{\bm{s}}(\rho_{\bm{m}})$. 
    
    \item We now aim to run a noiseless QEC program that measures the stabilisers $\bm{S}$ on the quantum computer. But again, due to noise, a noisy QEC program $\mathsf{Q}$ is run instead, obtaining $|\bm{S}|$ noisy measurement outputs $\bm{s}$. We immediately apply the destabilisers $D(\bm{s})$ to the state in the hardware.
    
    \item To complete the decoding, we then input the noisy syndrome $\bm{s}$ (hence the approximation) into the ideal decoder, which will then output the optimal correction program $\mathsf{C}_{\bm{m}, \bm{s}}^*$. We apply such a correction program in the hardware, completing the decoding of the prepared state.
\end{itemize}

In practice, this process can be pre-calculated for every measurement output $\bm{m}$ and noisy syndrome $\bm{s}$ into a CL-ML-LUT.
As long as the number of possible $\bm{m}$ and $\bm{s}$ is not too large, this look-up table can be stored in a fast memory close to the hardware, then executed with no algorithmic and negligible communication delay. 
If the number of possible measurement outputs is too large to be computed and/or stored, but their probability distribution has a manageable number of them that are more likely than the rest, one can run a Monte Carlo simulation to find the most likely ones, and then compute and store the ideal correction for those. 
We can additionally pipeline this ideal decoder with an algorithmic decoder, so that if a pair of $\bm{m}$ and $\bm{s}$ is observed on the hardware but is not in the look-up table, the algorithmic decoder can propose a correction.

\section{Example} 
\label{sec:example}
We now exemplify many of the definitions and methods described in the previous sections to show how natural the framework is for making complex calculations, even by hand. 
Our guiding example consists of a static program composed of several quantum operations to prepare and correct a 3-qubit repetition code that protects from $X$ errors.

The simulation starts by initialising four qubits $\mathcal{Q} = \{q_1, q_2, q_3, q_4\}$ in the state $\ket{0000}$, which contains the logical state $\ket{\Bar{0}}=\ket{000}$ on the first three qubits. 
Using the update rule~(\ref{box:projection}), the normalisation factor is initialised as $w=1$ and the tableau as
\begin{equation}
    \bm{T} = (Z_{q_1},\, Z_{q_2},\, Z_{q_3},\, Z_{q_4}).
\end{equation}
    
We now apply a $R^\theta_{X_{q_1}}$ rotation as a coherent noise channel. 
Using the update rule~(\ref{box:rotation}), the set of rotation qubits gets initialised as $\mathcal{R} = \{r_1\}$ and the tableau updates to
\begin{equation}
    \bm{T} = (Z_{q_1}C_{r_1},\, Z_{q_2},\, Z_{q_3},\, Z_{q_4},\, X_{q_1}O_{r_1}).
\end{equation}

The next step is to implement the Clifford gates $CX(q_1,q_4) \, CX(q_2,q_4)$ in order to measure the stabiliser operator $Z_{q_1}Z_{q_2}$ of the repetition code on the output of the ancilla $q_4$. 
Using the update rule~(\ref{box:clifford}), the tableau updates to 
\begin{equation}
    \bm{T} = (Z_{q_1}C_{r_1},\, Z_{q_2},\, Z_{q_3},\, Z_{q_1}Z_{q_2}Z_{q_4},\, X_{q_1}X_{q_4}O_{r_1}).
\end{equation}

Now apply three Pauli-flip channels $\mathcal{E}^{p_{f_1}}_{X_{q_1}}$, $\mathcal{E}^{p_{f_2}}_{X_{q_2}}$ and $\mathcal{E}^{p_{f_3}}_{X_{q_3}}$ as noise channels. Using the update rule~(\ref{box:pauli-flip}), the set of flip qubits gets initialised as $\mathcal{F}=\{f_1, f_2, f_3\}$, the normalisation weight updates to $w=1/8$, and the tableau updates to 
\begin{equation}
    \bm{T} = (Z_{q_1}C_{r_1}F_{f_1},\, Z_{q_2}F_{f_2},\, Z_{q_3}F_{f_3},\, Z_{q_1}Z_{q_2}Z_{q_4}F_{f_1}F_{f_2},\, X_{q_1}X_{q_4}O_{r_1}).
\end{equation}

Before continuing, let us replace the first element with the product of the first, second and fourth elements, and the fourth element with the product of the fourth and the second. This produces an equivalent tableau with all the rotation and flip operators decoupled, i.e., not present simultaneously in any tableau element. The tableau becomes
\begin{equation} \label{eq:bef_reas_anc}
    \bm{T} = (Z_{q_4}C_{r_1},\, Z_{q_2}F_{f_2},\, Z_{q_3}F_{f_3},\, Z_{q_1}Z_{q_4}F_{f_1},\, X_{q_1}X_{q_4}O_{r_1}).
\end{equation}
The tableau is now in a more convenient form to take the partial trace over the rotation and flip qubits.
This shows that the tableau represents the expected (in this case normalised) state on the qubit space:
\begin{equation}
\begin{split}
    \rho|_\mathcal{Q} & = \tfrac{1}{8} \Tr_{r_1}\Pi(Z_{q_4}C_{r_1},\, X_{q_1}X_{q_4}O_{r_1})\Tr_{f_1}\Pi(Z_{q_1}Z_{q_4}F_{f_1})\Tr_{f_2}\Pi(Z_{q_2}F_{f_2})\Tr_{f_3}\Pi(Z_{q_3}F_{f_3})\\
    &= \tfrac{1}{16}(I+\cos{\theta_{r_1}}Z_{q_4} - \sin{\theta_{r_1}}X_{q_1}Y_{q_4})(I+\epsilon_{f_1}Z_{q_1}Z_{q_4})(I+\epsilon_{f_2}Z_{q_2})(I+\epsilon_{f_3}Z_{q_3}).
\end{split}
\end{equation}
Notice how the computation naturally factorises into the computation of four independent traces. 
The discovery of factorisation strategies can significantly reduce the exponential cost of computing with SyQMA.

We will now add a measurement to detect any problematic errors.
First, measure the ancilla $q_4$ in the Pauli basis $\bm{P} = (Z_{q_4})$ to detect problematic faults. 
To do it, first project onto the measurement output $\bm{m}=(m_{q_4})$ using the update rule~(\ref{box:projection}), which updates the normalisation weight to $w=1/16$ and the tableau to
\begin{equation}
    \bm{T} = (Z_{q_4}C_{r_1},\, Z_{q_2}F_{f_2},\, Z_{q_3}F_{f_3},\, Z_{q_1}Z_{q_4}F_{f_1},\, m_{q_4}Z_{q_4}).
\end{equation}
Second, trace out the ancilla $q_4$ using the update rule~(\ref{box:trace-out}), which updates the qubit set to $\mathcal{Q}=\{q_1, q_2, q_3\}$ and the tableau updates to
\begin{equation}
    \bm{T} = (m_{q_4}C_{r_1},\, Z_{q_2}F_{f_2},\, Z_{q_3}F_{f_3},\, m_{q_4}Z_{q_1}F_{f_1}).
\end{equation}

The goal was to implement a noiseless program with noiseless circuit branches, but due to the noise channels labelled by the rotation qubit $r_1$ and the Pauli-flip qubits $f_1, f_2, f_3$, we have actually implemented a noisy quantum program $\mathsf{P}$ that has led to the noisy circuit branches $\mathsf{P}_{\bm{m}}$ and the noisy state $\rho_{\bm{m}}=\mathsf{P}_{\bm{m}}()$ for the measurement output $\bm{m}=(m_{q_4})$.
\\

We now continue the example with a QEC analysis.
We start by directly computing (without previous decoding) the expectation value of the logical operator $\Bar{Z}=Z_{q_2}$ and the logical success rate (LSR) from Eq.~\eqref{eq:average_lsr}, i.e., the probability of measuring it in $+1$ using the computation of the trace~(\ref{box:trace}):
\begin{align}
    \Tr(\rho_{\bm{m}}) &= (1 + m_{q_4}\cos{\theta_{r_1}})/2, \\
    \Tr(\Bar{Z}\rho_{\bm{m}}) &= (\epsilon_{f_2} + m_{q_4}\epsilon_{f_2}\cos{\theta_{r_1}})/2, \\ 
    \mathbb{E}[\Bar{Z}|\rho_{\bm{m}}] &= \Tr(\Bar{Z}\rho_{\bm{m}}) / \Tr(\rho_{\bm{m}}) =  \epsilon_{f_2} = 1-2p_{f_2},\\
    \mathrm{LSR}[\rho_{\bm{m}}] &=\mathbb{P}[\Bar{Z}=+1|\rho_{\bm{m}}] = \Tr\left(\Pi(\Bar{Z}) \rho_{\bm{m}}\right) / \Tr(\rho_{\bm{m}}) = (1+\epsilon_{f_2})/2 = 1-p_{f_2}.
\end{align}
So the Pauli-flip channel $\mathcal{E}^{p_{f_2}}_{X_{q_2}}$ is the only one that affects the expectation value of the logical operator and the probability of measuring it in $+1$.
The LER is then $p_{f_2}$, meaning that measuring the logical operator at this point, without decoding it, would lead to a $p_{f_2} \in \mathcal{O}(p)$ LER.

To decode the state in the simulator, we start by implementing the decoding program $\mathsf{D}$, with branches $\mathsf{D}_{\bm{s}}$.
This program first projects onto the syndrome $\bm{s}\in\{\pm1\}^2$ for the stabilisers $\bm{S}=(Z_{q_1}Z_{q_2},\, Z_{q_2}Z_{q_3})$. 
The tableau updates to
\begin{equation}
    \bm{T} = (m_{q_4}C_{r_1},\, Z_{q_2}F_{f_2},\, Z_{q_3}F_{f_3},\, m_{q_4}Z_{q_1}F_{f_1},\, s_1Z_{q_1}Z_{q_2},\, s_2Z_{q_2}Z_{q_3}).
\end{equation}
The decoding program then applies the product $D(\bm{s})=\prod_{i=1}^{|\bm{s}|}D_i^{(1-s_i)/2}$.
For our choice of logical operators $\Bar{Z}=Z_{q_2}$ and $\Bar{X}=X_{q_1}X_{q_2}X_{q_3}$,
the destabilisers are $D_i \in \bm{D}=(X_{q_1},\, X_{q_3})$. 
This just multiplies by $s_1$ ($s_2$) every element that anticommutes with $D_1$ ($D_2$). 
The tableau that represents the noisy state $\mathsf{D}_{\bm{s}}(\title{\rho}_{\bm{m}})$ therefore becomes
\begin{equation}
    \bm{T} = (m_{q_4}C_{r_1},\, Z_{q_2}F_{f_2},\, s_2Z_{q_3}F_{f_3},\, s_1m_{q_4}Z_{q_1}F_{f_1},\, Z_{q_1}Z_{q_2},\, Z_{q_2}Z_{q_3}).
\end{equation}
Note that the represented state is stabilised in $+1$ by the two stabilisers (the last two elements). This means the state in the code space, so all the noise has become a logical error channel.

The logical stabiliser is $\Bar{L}=\Bar{Z}$ and we consider the correction programs $\mathsf{C}^{(+1)}(\rho)=\rho$ and $\mathsf{C}^{(-1)}(\rho) = \Bar{X}\rho \Bar{X}$ for any input state $\rho$. 
The update on the tableau under this correction program $\mathsf{C}^{(x)}$ parameterised by $x\in\{\pm1\}$ consists in multiplying every element that anticommutes with the logical operator $\Bar{X}$ by $x$:
\begin{equation}
    \bm{T} = (m_{q_4}C_{r_1},\, xZ_{q_2}F_{f_2},\, xs_2Z_{q_3}F_{f_3},\, xs_1m_{q_4}Z_{q_1}F_{f_1},\, Z_{q_1}Z_{q_2},\, Z_{q_2}Z_{q_3}).
\end{equation}

We get the canonical form $\bm{T} = (\bm{T}_{\mathcal{Q}}, \bm{T}_{\mathcal{R}}, \bm{T}_{\mathcal{F}}, \bm{T}_{\emptyset})$ for the state with the parameterised correction program applied on it $\mathsf{C}^{(x)}\circ\mathsf{D}_{\bm{s}}(\rho_{\bm{m}})$:
\begin{align}
    \bm{T}_{\mathcal{Q}} &= (xs_1m_{q_4}Z_{q_1}F_{f_1},\, xZ_{q_2}F_{f_2},\, xs_2Z_{q_3}F_{f_3}), \\
    \bm{T}_{\mathcal{R}} &= (m_{q_4}C_{r_1}), \\
    \bm{T}_{\mathcal{F}} &= (s_1 m_{q_4}F_{f_1}F_{f_2},\, s_2F_{f_2}F_{f_3}), \\
    \bm{T}_{\emptyset} &= \emptyset.
\end{align}
The first element in $\bm{T}_{\mathcal{F}}$ is the product of the second, fourth and fifth elements in the tableau, while the second element in $\bm{T}_{\mathcal{F}}$ is the product of the second, third, and sixth.
Since $\bm{T}_{\emptyset} = \emptyset$, the set of constraints is $\mathcal{C}=\emptyset$, which is trivially satisfied for all values of the circuit parameters.

The measurement probability of $(m_{q_4}, s_1, s_2)$ is the transition probability from the state $\rho$ immediately before the first Pauli measurement (represented by the tableau in Eq.~\eqref{eq:bef_reas_anc}) to the state $\mathsf{C}^{(x)}\circ\mathsf{D}_{\bm{s}}(\rho_{\bm{m}})$, represented by the canonical form above:
\begin{equation}
    \mathbb{P}[\mathsf{C}^{(x)}\circ\mathsf{D}_{\bm{s}}(\rho_{\bm{m}})|\rho] = (1+m_{q_4}\cos{\theta_{r_1}})(1+s_1 m_{q_4}\epsilon_{f_1}\epsilon_{f_2} + s_2\epsilon_{f_2}\epsilon_{f_3} + s_1s_2 m_{q_4}\epsilon_{f_1}\epsilon_{f_3})/8.
\end{equation}

To decide the optimal correction program for every circuit branch and syndrome, we compute the LSR, i.e., the probability of measuring the logical stabiliser in $\Bar{L}=+1$ for every branch and syndrome:
\begin{align}
    \mathrm{LSR}[\mathsf{C}^{(x)}\circ\mathsf{D}_{\bm{s}}(\rho_{\bm{m}})] &= \mathbb{P}[\Bar{L}=+1 | \mathsf{C}^{(x)}\circ\mathsf{D}_{\bm{s}}(\rho_{\bm{m}})] = \tfrac{1}{2}\left(1 + \mathbb{E}[\Bar{L} | \mathsf{C}^{(x)}\circ\mathsf{D}_{\bm{s}}(\rho_{\bm{m}})]\right),\\
    \mathbb{E}[\Bar{L} | \mathsf{C}^{(x)}\circ\mathsf{D}_{\bm{s}}(\rho_{\bm{m}})] &= x\,\mathbb{E}[\Bar{L} | \mathsf{D}_{\bm{s}}(\rho_{\bm{m}})] = x\,\frac{\epsilon_{f_2}+s_1 m_{q_4}\epsilon_{f_1} + s_2\epsilon_{f_3} + s_1s_2 m_{q_4}\epsilon_{f_1}\epsilon_{f_2}\epsilon_{f_3}}{1+s_1 m_{q_4}\epsilon_{f_1}\epsilon_{f_2} + s_2\epsilon_{f_2}\epsilon_{f_3} + s_1s_2 m_{q_4}\epsilon_{f_1}\epsilon_{f_3}}.
\end{align}

The optimal correction program $\mathsf{C}_{\bm{m},\bm{s}}^* = \mathsf{C}^{(x^*_{\bm{m},\bm{s}})}$ is therefore given by
\begin{equation}
    x^*_{\bm{m},\bm{s}}=\mathrm{sign}\,\mathbb{E}[\Bar{L} | \mathsf{D}_{\bm{s}}(\rho_{\bm{m}})]=\mathrm{sign}(\epsilon_{f_2}+s_1 m_{q_4}\epsilon_{f_1} + s_2\epsilon_{f_3} + s_1s_2 m_{q_4}\epsilon_{f_1}\epsilon_{f_2}\epsilon_{f_3}).
\end{equation}
In the second equality, we have used the fact that the denominator in the expectation value is positive semi-definite (because it is the trace of a state), so it does not affect the sign of the expectation value.
We then obtain the dynamic correction map $\mathsf{C}^*$ that implements the static program $\mathsf{C}_{\bm{m},\bm{s}}^*$ for every branch and syndrome.

The average LSR over every circuit branch $\bm{m}=(m_{q_4})$ and syndrome $\bm{s}=(s_1,s_2)$ when running the noisy state preparation program $\mathsf{P}$ followed by the obtained optimal correction program $\mathsf{C}^*$ is therefore obtained by taking the absolute value of the expectation value:
\begin{equation}
    \mathrm{LSR}[\mathsf{C}^*\circ\mathsf{D}\circ\mathsf{P}] = \tfrac{1}{2}\sum_{\bm{m}}\sum_{\bm{s}}\left(1 + \left|\mathbb{E}[\Bar{L} | \mathsf{D}_{\bm{s}}(\rho_{\bm{m}})]\right|\right) \, \mathbb{P}[\mathsf{D}_{\bm{s}}(\rho_{\bm{m}})|\rho] =
    \tfrac{1}{2} + \tfrac{1}{2}\sum_{\bm{m}}\sum_{\bm{s}}\left| \Tr\left(\Bar{L}\, \mathsf{D}_{\bm{s}}(\rho_{\bm{m}})\right)\right|.
\end{equation}

To continue with this simple example, we assume that all noise events are equally likely: $\epsilon_{f_1}= \epsilon_{f_2}=\epsilon_{f_3}=\epsilon = 1-2p \in [0, 1]$, with $p\in[0,1/2]$. Then, the optimal correction program has $x^*=-1$ for the branches and syndromes $m_{q_4}=+1, \,s_1=s_2=-1, $ and $m_{q_4}=s_1=+1, \, s_2=-1$, i.e., for these, the decoder applies the logical $\Bar{X}$ Pauli operator as the optimal correction program. For the rest of branches and syndromes the correction program leaves the state untouched. 
The average LSR is then
\begin{equation}
    \mathrm{LSR}[\mathsf{C}^*\circ\mathsf{D}\circ\mathsf{P}] = \tfrac{1}{2}\left(1+\tfrac{1}{4}|3\epsilon+\epsilon^3| + \tfrac{3}{4}|\epsilon-\epsilon^3|\right) = 1-3p^2 + 2p^3.
\end{equation}
So, as expected for a distance-$3$ code, the LER $3p^2 - 2p^3$ of a fault-tolerant preparation circuit followed by an ideal decoder is $\mathcal{O}(p^2)$.

\section{Results} \label{sec:results}
This section presents the simulation and analysis of fault-tolerant (FT) stabiliser and magic state preparations on various small quantum error correcting (QEC) codes.
We start with the set-up of the code definitions, noise model and state preparation program, and then describe the expectation values (EVs) and logical error rates (LERs) evaluated on them. In the following subsections, we report LERs for several codes and highlight two additional analyses: in Section~\ref{sec:decoding_analysis}, we investigate the uncertainty of correctly decoding each syndrome and the improvements gained through postselection, and in Section~\ref{sec:memory_time}, we prove our claims that SyQMA can only consume a polynomial amount of memory to calculate EVs and LERs.

\subsection{Introduction to QEC codes}
Following Section~\ref{sec:qec}, an $\llbracket n,k,d\rrbracket$ QEC code encodes $k$ logical qubits in $n$ physical qubits at (odd) distance $d$, i.e., $t+1$ or more faults (with $t = (d-1)/2$) can potentially produce a logical error. 
We consider the fault-tolerant preparation of the Steane code $\llbracket7,1,3\rrbracket$ in the stabiliser state $\ket{\Bar{0}}$ and the magic state $\ket{\Bar{H}}=\Bar{R}^{\pi/4}_{\Bar{Y}}\ket{\Bar{0}}$, the 3D colour code $\llbracket15,1,3\rrbracket$ prepared in the stabiliser state $\ket{\Bar{0}}$ and the magic state $\ket{\Bar{T}}=\Bar{R}^{\pi/4}_{\Bar{Z}}\ket{\Bar{+}}$, and the 2D colour code $\llbracket17,1,5\rrbracket$ prepared in the $\ket{\Bar{0}}$ state.
The codes have stabiliser generators $\bm{S}$ and destabilisers $\bm{D}$.
All these codes are CSS~\cite{calderbank1996good}, i.e., all the stabilisers and logical operators of the code are multi-qubit Pauli operators composed only of $X$ or only of $Z$ Pauli strings, which allows the independent correction of $X$ and $Z$ errors.
The states prepared $\ket{\Bar{\psi}}$ are stabilised by the logical stabiliser $\Bar{L}$, which is $\Bar{Z}$ for $\ket{\Bar{0}}$, $(\Bar{Z}+\Bar{X})/\sqrt{2}$ for $\ket{\Bar{H}}$ and $(\Bar{X}+\Bar{Y})/\sqrt{2}$ for $\ket{\Bar{T}}$.

We use a typical circuit-level noise model in QEC, with single-qubit depolarising channels after every single-qubit gate, and two-qubit depolarising channels after every two-qubit gate, all of them with the same unified error rate $p$.
Under this noise model every component fails independently with an $O(p)$ probability, two faults happen with probability $O(p^2)$, and so on.
The plotted discard rates and LERs are analytical and exact functions of the physical error rate $p$, i.e., no approximation is applied, not even due to Monte Carlo sampling of the noise.

The noisy quantum program $\mathsf{P}$ that prepares the logical state implements a quantum circuit that contains flag and ancilla qubits to detect errors and discards the circuit upon their detection.
We therefore only retain the noisy state $\rho = \mathsf{P}_{\bm{+1}}()$ obtained by the state preparation program branch where the measurement output is a string of $+1$'s, indicating that no error has been detected. 
The acceptance rate is the probability $\mathbb{P}[\rho] = \Tr(\rho)$ of not detecting any error and the discard rate $1 - \mathbb{P}[\rho]$ is the probability of detecting any error.
Since there are single faults (events with $O(p)$ probability) that are detected, the discard rate is expected to scale as $O(p)$.

\subsection{Expectation values and logical error rates}
On the accepted states we compute the following \textit{uncorrected} expectation values and LERs directly, i.e., without previous decoding:
\begin{align}
    \mathbb{E}[\Bar{O}|\rho] &= \Tr\left(\Bar{O}\rho\right) \, / \, \Tr(\rho), \label{eq:uncorr_ev} \\
    \mathrm{LER}[\Bar{O} | \rho] &= \tfrac{1}{2}\bra{\Bar{\psi}}\Bar{O}\ket{\Bar{\psi}} - \tfrac{1}{2}\mathbb{E}[\Bar{O}|\rho], \label{eq:uncorr_lerO}\\
    \mathrm{LER}[\rho] &= 1 - \mathrm{LSR}[\rho] = \tfrac{1}{2} - \tfrac{1}{2}\mathbb{E}[\Bar{L}|\rho], \label{eq:uncorr_ler}
\end{align}
for the logical operators $\Bar{O} \in \{\Bar{X}, \, \Bar{Z}\}$.
The second equation is the probability of errors anticommuting with the logical operator $\Bar{O}$ if we assume that the noisy state $\rho$ is equal to the noiseless state $\ket{\Bar{\psi}}$ up to Pauli errors.
Since no decoding has been performed, the LERs are expected to be $O(p)$. 

For decoding, we project onto the syndrome $\bm{s}$ of the stabilisers and apply the corresponding destabilisers, obtaining the state $\mathsf{D}_{\bm{s}}(\rho) = D(\bm{s}) \, \Pi(\bm{s}\bm{S}) \, \rho \, \Pi(\bm{s}\bm{S}) \, D(\bm{s})$.
This state is in the code space, but it might carry a logical error channel.
We consider correction programs $\mathsf{C}^{(x,z)}$ that apply logical operators $\Bar{P}(x,z)$ to the state.
Under these correction programs, the expectation values of the logical operators change sign depending on the sign parameters $(x,z)$ as $\mathbb{E}[\Bar{X}|\mathsf{C}^{(x,z)}\circ\mathsf{D}_{\bm{s}}(\rho)] = z \, \mathbb{E}[\Bar{X}|\mathsf{D}_{\bm{s}}(\rho)]$ and $\mathbb{E}[\Bar{Z}|\mathsf{C}^{(x,z)}\circ\mathsf{D}_{\bm{s}}(\rho)] = x \, \mathbb{E}[\Bar{Z}|\mathsf{D}_{\bm{s}}(\rho)]$.

The optimal correction programs $\mathsf{C}^*_{\bm{s}} = \mathsf{C}^{(x^*_{\bm{s}},z^*_{\bm{s}})}$ are those that maximise the logical success rate $\mathrm{LSR}[\mathsf{P}] = \mathbb{P}[\Bar{L}=+1 |\mathsf{C}^{(x,z)}\circ\mathsf{D}_{\bm{s}}(\rho)] = \tfrac{1}{2} + \tfrac{1}{2}\mathbb{E}[\Bar{L}|\mathsf{C}^{(x,z)}\circ\mathsf{D}_{\bm{s}}(\rho)]$ for each syndrome $\bm{s}$, i.e., the probability of measuring the logical stabiliser $\Bar{L}$ in $+1$ on the states $\mathsf{D}_{\bm{s}}(\rho)$.
Therefore, for all states considered in this section, the optimal correction programs are defined by the binary functions $x^*_{\bm{s}} = \mathrm{sign}\,\mathbb{E}[\Bar{Z}|\mathsf{D}_{\bm{s}}(\rho\textit{})]$ and $z^*_{\bm{s}} = \mathrm{sign}\,\mathbb{E}[\Bar{X}|\mathsf{D}_{\bm{s}}(\rho)]$.
This solution leads to the \textit{corrected} expectation values and LERs:
\begin{align}
    \mathbb{E}[\Bar{O}|\mathsf{C}^*_{\bm{s}}\circ\mathsf{D}_{\bm{s}}(\rho)] &= \left|\mathbb{E}[\Bar{O}|\mathsf{D}_{\bm{s}}(\rho)]\right|, \label{eq:corr_ev}\\
    \mathrm{LER}[\Bar{O} | \mathsf{C}^*_{\bm{s}}\circ\mathsf{D}_{\bm{s}}(\rho)] &= \tfrac{1}{2}\bra{\Bar{\psi}}\Bar{O}\ket{\Bar{\psi}} - \tfrac{1}{2}\mathbb{E}[\Bar{O}|\mathsf{C}^*_{\bm{s}}\circ\mathsf{D}_{\bm{s}}(\rho)]. \label{eq:corr_lerO}\\
    \mathrm{LER}[\mathsf{C}^*_{\bm{s}}\circ\mathsf{D}_{\bm{s}}(\rho)] &= 1 - \mathrm{LSR}[\mathsf{C}^*_{\bm{s}}\circ\mathsf{D}_{\bm{s}}(\rho)] = \tfrac{1}{2} - \tfrac{1}{2}\mathbb{E}[\Bar{L}|\mathsf{C}^*_{\bm{s}}\circ\mathsf{D}_{\bm{s}}(\rho)]. \label{eq:corr_ler}
\end{align}
The second equation is the probability that the logical error anticommutes with the logical operator $\Bar{O}$, assuming that the noisy state $\mathsf{D}_{\bm{s}}(\rho)$ is equal to the noiseless state $\ket{\Bar{\psi}}$ up to logical Pauli errors. 
Thanks to decoding, the LERs are expected to improve to $O(p^{t+1})$. 
As explained in Section~\ref{sec:qec}, this corresponds to conducting circuit-level maximum likelihood (CL-ML) decoding. 
We also report the optimal correction programs as CL-ML look-up tables (CL-ML-LUTs), together with additional soft information that can be used for decoding. 

We finally consider retaining only the states $\mathsf{D}_{\bm{+1}}(\rho)$ of trivial syndrome $\bm{s}=\bm{+1}$ and apply no correction program. 
This results in the \textit{postselected} expectation values and LERs:
\begin{align}
    \mathbb{E}[\Bar{O}|\mathsf{D}_{\bm{+1}}(\rho)] &= \left|\mathbb{E}[\Bar{O}|\mathsf{D}_{\bm{+1}}(\rho)]\right|, \label{eq:post_ev}\\
    \mathrm{LER}[\Bar{O} | \mathsf{D}_{\bm{+1}}(\rho)] &= \tfrac{1}{2}\bra{\Bar{\psi}}\Bar{O}\ket{\Bar{\psi}} - \tfrac{1}{2}\mathbb{E}[\Bar{O}|\mathsf{D}_{\bm{+1}}(\rho)]. \label{eq:post_lerO}\\
    \mathrm{LER}[\mathsf{D}_{\bm{+1}}(\rho)] &= 1 - \mathrm{LSR}[\mathsf{D}_{\bm{+1}}(\rho)] = \tfrac{1}{2} - \tfrac{1}{2}\mathbb{E}[\Bar{L}|\mathsf{D}_{\bm{+1}}(\rho)]. \label{eq:post_ler}
\end{align}
The second equation is the probability that the logical error anticommutes with the logical operator $\Bar{O}$, assuming that the noisy state $\mathsf{D}_{\bm{+1}}(\rho)$ is equal to the noiseless state $\ket{\Bar{\psi}}$ up to logical Pauli errors. 
Thanks to postselecting, the LERs are expected to be between $O(p^{t+1})$ and $O(p^d)$. 
For the three decoding strategies: uncorrected, corrected, and postselected LERs, the second equation reduces to the third equation (obtained from Eq.~\eqref{eq:average_lsr}) when evaluating the LER of the logical stabiliser, i.e., when $\Bar{O}=\Bar{L}$.

In the current version of our package, we can only display analytical expressions for uncorrected and postselected EVs/LERs, which we provide in the captions of the figures where memory allows us to store the full expressions. However, internally, all LERs are calculated symbolically and without approximation. For brevity, in the captions we display only the first few significant digits of the coefficients of the leading order physical noise terms, but the software that we provide calculates the coefficients to full precision. In all the legends, we use numerical fits of the LERs for the lowest physical error rates, as this does not require storing entire symbolic expressions.

We plan to expand these capabilities in a future version to also obtain expressions for discard and corrected LERs and also to avoid the requirement of storing the full expressions.

\subsection{$\llbracket k+2,k,2\rrbracket$ Iceberg code}
The Iceberg code~\cite{self2024protecting, grassl2004optimal} is a $\llbracket k+2,k,2\rrbracket$ code that encodes $k$ (even) logical qubits into $n = k + 2$ physical qubits and can detect (but not correct) any single-fault. 
This minimum overhead, together with the native implementation of partially-FT arbitrary-angle Pauli rotations, makes the Iceberg code very useful to protect near-term quantum circuits in today's quantum computers~\cite{yamamoto2024demonstrating, jin2025iceberg, he2025performance}.
The two stabilisers of the code have support on all the qubits: $\bm{S} = (Z^{\otimes n},\, X^{\otimes n})$, and the logical operators are $\Bar{X}_i = X_iX_t$ and $\Bar{Z}_i = Z_iZ_b$ for all $i=[k]$, where $t$ and $b$ are the indices of the two additional code qubits.

\begin{figure}[htbp]
    \centering
    \begin{minipage}[b]{0.48\textwidth}
        \centering
        
        \begin{subfigure}[b]{0.52\linewidth}
            \centering
            \resizebox{\linewidth}{!}{
            \begin{tikzpicture}
                \begin{yquant}
                    qubit {$\ket{\makebox[\slength][c]{0}}$} q[8];
                    qubit {$\ket{\makebox[\slength][c]{0}}$} a[1];

                    h q[0];
                    cnot q[1] | q[0];
                    cnot q[2] | q[1];
                    cnot q[3] | q[2];
                    cnot q[4] | q[3];
                    cnot q[5] | q[4];
                    cnot q[6] | q[5];
                    cnot q[7] | q[6];
                    cnot a[0] | q[0];
                    cnot a[0] | q[7];

                    measure a[0];
                \end{yquant}
            \end{tikzpicture}}
            \caption{}
            \label{fig:iceberg_1_branch}
        \end{subfigure}
        \hfill
        \begin{subfigure}[b]{0.46\linewidth}
            \centering
            \resizebox{\linewidth}{!}{
            \begin{tikzpicture}
                \begin{yquant}
                    qubit {$\ket{\makebox[\slength][c]{0}}$} q[8];
                    qubit {$\ket{\makebox[\slength][c]{0}}$} a[2];

                    h q[0];
                    cnot q[4] | q[0];
                    cnot q[2] | q[0];
                    cnot q[6] | q[4];

                    cnot q[1] | q[0];
                    cnot q[3] | q[2];
                    cnot q[5] | q[4];
                    cnot q[7] | q[6];

                    cnot a[0] | q[7];
                    cnot a[1] | q[5];
                    cnot a[0] | q[3];
                    cnot a[1] | q[1];

                    measure a[0];
                    measure a[1];
                \end{yquant}
            \end{tikzpicture}}
            \caption{}
            \label{fig:iceberg_4_branch}
        \end{subfigure}

        \vspace{0.5cm}

        \begin{subfigure}[b]{\linewidth}
            \centering
            \resizebox{\linewidth}{!}{
            \begin{tikzpicture}
                \begin{yquant}
                    qubit {$\ket{\makebox[\slength][c]{0}}$} q[8];
                    qubit {$\ket{\makebox[\slength][c]{0}}$} a[2];

                    h a[1];
                    cnot q[0] | a[1];
                    cnot a[0] | q[0];
                    cnot a[0] | q[1];
                    cnot q[1] | a[1];

                    barrier (-);

                    cnot q[2] | a[1];
                    cnot a[0] | q[2];
                    cnot q[3] | a[1];
                    cnot a[0] | q[3];

                    barrier (-);
                    
                    cnot q[4] | a[1];
                    cnot a[0] | q[4];
                    cnot q[5] | a[1];
                    cnot a[0] | q[5];

                    barrier (-);

                    cnot q[6] | a[1];
                    cnot a[0] | q[6];
                    cnot a[0] | q[7];
                    cnot q[7] | a[1];
                    h a[1];

                    align q, a;
                    measure a[0];
                    measure a[1];
                \end{yquant}
            \end{tikzpicture}}
            \caption{}
            \label{fig:iceberg_synd_meas}
        \end{subfigure}
    \end{minipage}
    \hfill
    \begin{minipage}[b]{0.5\textwidth}
        \centering
        \begin{subfigure}[b]{\linewidth}
            \centering
            \includegraphics[width=1.0\linewidth]{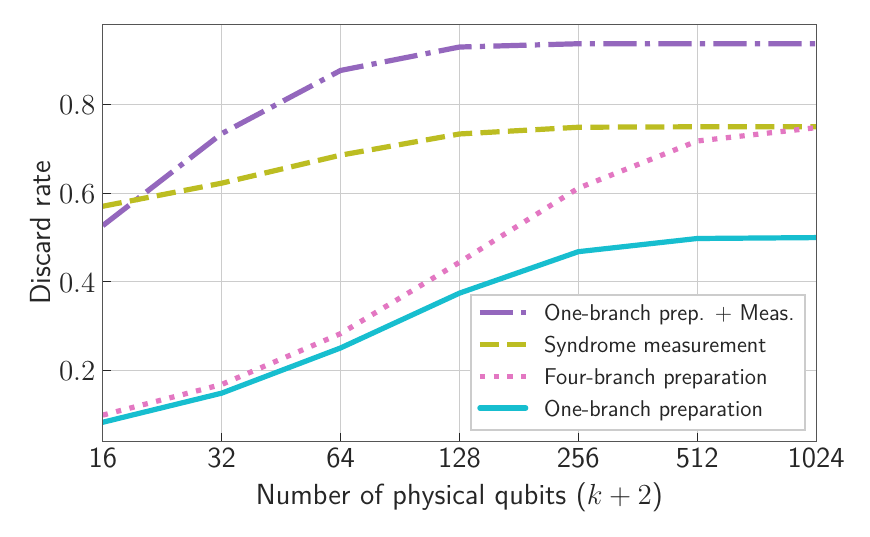}
            \caption{}
            \label{fig:iceberg_discard}
        \end{subfigure}
    \end{minipage}

    \caption{Iceberg code fault-tolerant state preparations and syndrome measurement. (a) One-branch state preparation, (b) four-branch state preparation, and (c) syndrome measurement for the $\llbracket8,6,2\rrbracket$ code. The first and last blocks in the syndrome measurement circuit are the same for every $k$, but the middle of the circuit consists of a block repeated $k/2-1$ times (twice in this example). (d) Discard rates for different state preparation and syndrome measurements gadgets in the $\llbracket k+2,k,2\rrbracket$ Iceberg code. The physical error rate is fixed at $p = 10^{-2}$.}
    \label{fig:iceberg_circuits}
\end{figure}

The circuits for the $\ket{\Bar{0}}^{\otimes k}$ state preparations and syndrome measurement in the Iceberg code are presented in Fig.~\ref{fig:iceberg_circuits} \cite{self2024protecting, dasu2026computing}. Figs.~\ref{fig:iceberg_1_branch} and~\ref{fig:iceberg_4_branch} have the structure of one branch, which uses one ancilla, and the structure of four branches, which uses two ancillas but lower depth, respectively.
The state preparation program $\mathsf{P}$ runs one of these circuits, tracing out and restarting all qubits until the ancillas produce the measurement output $\bm{+1}$, corresponding to the circuit branch $\mathsf{P}_{\bm{+1}}$ and the unique accepted state $\rho=\mathsf{P}_{\bm{+1}}()$.
Fig.~\ref{fig:iceberg_discard} shows the discard rates $1-\Tr(\rho)$, i.e., the probability that at least one of the ancillas detects an error, for the two state preparation circuits and the FT syndrome measurement, assuming no noise before their execution.

In a $d = 2$ distance code, single faults (errors inserted with probability $O(p)$) can be detected by the ancilla measurements. 
The figure shows the discard rate as a function of the number $n$ of data qubits, for a fixed value of the physical error rate $p = 10^{-2}$. Each ancilla measurement is deterministic in the absence of noise, but becomes non-deterministic when noise is included.
As we discussed in Section~\ref{sec:complexity}, the runtime of our simulator increases exponentially with the number of these measurements. 
But in these circuits, since there are only one or two such measurements, the runtime increases only by a factor of two or four over the polynomial cost of traditional Clifford simulation.
Since the number of ancillas in the circuit is independent of the system size $n$, the simulation can grow to such large system sizes as those presented in Fig.~\ref{fig:iceberg_discard}.
Note that, as the code size grows, the discard rate approaches the limit of the maximally mixed state of $1 - 2^{\mathrm{-\#ancillas}}$, due to the increasing noise accumulated.

\subsection{$\llbracket7,1,3\rrbracket$ Steane code}
The Steane colour code is a self-dual code, i.e., the transversal physical Hadamard $H^{\otimes 7}$ preserves the stabiliser group of the code and implements a logical Hadamard $\Bar{H}$.
The self-duality can be employed to fault-tolerantly prepare magic states with zero-level distillation~\cite{goto2016minimizing, chamberland2019fault}.

\subsubsection{Fault-tolerant Clifford state preparation}

\begin{figure}[h]
    \centering
    \begin{subfigure}[c]{0.49\textwidth}
        \centering
        \resizebox{\linewidth}{!}{
        \begin{tikzpicture}
            \begin{yquant}
                qubit {$\ket{+}$} q[1];
                qubit {$\ket{\makebox[\slength][c]{0}}$} q[+3];
                qubit {$\ket{+}$} q[+1];
                qubit {$\ket{\makebox[\slength][c]{0}}$} q[+1];
                qubit {$\ket{+}$} q[+1];
        
                cnot q[1] | q[0];
                cnot q[5] | q[4];
                cnot q[3] | q[6];
                cnot q[2] | q[4];
                cnot q[5] | q[6];
                cnot q[3] | q[0];
                cnot q[1] | q[4];
                cnot q[2] | q[3];

                align q[0], q[1], q[2], q[3], q[4], q[5], q[6];
                
                [after=q]
                qubit {$\ket{\makebox[\slength][c]{0}}$} z_anc[1];
                cnot z_anc[0] | q[1];
                cnot z_anc[0] | q[3];
                cnot z_anc[0] | q[5];
                
                measure z_anc[0];
            \end{yquant}
        \end{tikzpicture}}
        \caption{}
        \label{fig:circuit_steane}
    \end{subfigure}
    \hfill
    \begin{subfigure}[c]{0.49\textwidth}
        \centering
        \includegraphics[width=1.0\linewidth]{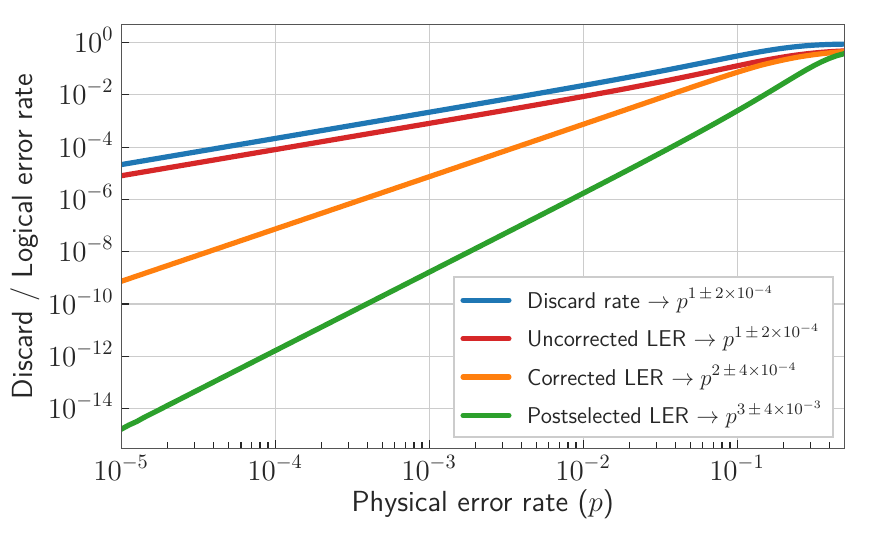}
        \caption{}
        \label{fig:steane_ft_ler_z}
    \end{subfigure}
    \caption{Fault-tolerant preparation of $\ket{\Bar{0}}$ in the Steane code. (a) Circuit for the FT state preparation of the logical state $\ket{\Bar{0}}$ in the Steane code. 
    (b) Discard rate and logical error rates under the three decoding strategies: uncorrected (Eq.~\eqref{eq:uncorr_ler}), corrected (Eq.~\eqref{eq:corr_ler}), and postselected (Eq.~\eqref{eq:post_ler}). The uncorrected LER is $0.8 \, p^1 + 11.8 \, p^2 + O(p^3)$ and the postselected LER is $1.63 \, p^3 + 42.98 \, p^4 + O(p^5)$.
    }
    \label{fig:placeholder}
\end{figure}

The circuit for the $\ket{\Bar{0}}$ FT state preparation in the Steane code is presented in Fig.~\ref{fig:circuit_steane}. Similar to the FT initialisation of the Iceberg code, the FT initialisation of the Steane code consists of a non-FT circuit (the first eight CNOTs) that prepares the desired state $\ket{\Bar{0}}$, followed by a verification circuit (the last three CNOTs and the ancilla) that detects any problematic fault~\cite{goto2016minimizing, ryan2021realization}.
The verification circuit detects any single fault (error event of $O(p)$ probability) that propagates to an error of minimum weight (up to multiplication by the stabilisers and $\Bar{L}$) larger than weight 1.
The state preparation program $\mathsf{P}$ runs this circuit, tracing out all qubits and restarting the circuit until the ancilla is measured in $+1$, leading to the circuit branch $\mathsf{P}_{+1}$ and the unique accepted state $\rho=\mathsf{P}_{+1}()$.

Fig.~\ref{fig:steane_ft_ler_z} shows that in the low physical error regime, the expected asymptotic scaling of the LERs for a $d = 3$ code with each of the strategies is recovered, with the actual symbolic expressions provided in the caption.
The legend of Fig.~\ref{fig:steane_ft_ler_z} shows the leading order of the polynomial fits to each of the LER: $O(p)$ for the uncorrected strategy (no decoding), $O(p^2)$ for the corrected strategy (CL-ML decoding), and $O(p^3)$ for the postselected strategy (accept only the trivial syndrome). 
The discard rate is $O(p)$, as expected from a FT preparation that discards the circuit on the detection of any error. 

The figure showcases SyQMA's capability to obtain LERs as low as $10^{-15}$ exactly and as cheaply in the low-error regime as in the high-error regime, with none of the usual sampling overhead.

\subsubsection{Fault-tolerant magic state preparation}
\begin{figure}[h]
    \centering
    \begin{minipage}[b]{0.49\textwidth}
        \centering
        \begin{subfigure}[b]{0.49\linewidth}
            \centering
            \resizebox{\linewidth}{!}{
            \begin{tikzpicture}
                \begin{yquant}
                    qubit {$\ket{+}$} q[2];
                    qubit {$\ket{\makebox[\slength][c]{$H$}}$} q[+1];
                    qubit {$\ket{+}$} q[+1];
                    qubit {$\ket{\makebox[\slength][c]{0}}$} q[+3];
                  
                    cnot q[4] | q[2];
                    cnot q[6] | q[0];
                    cnot q[5] | q[3];
                    
                    cnot q[5] | q[2];
                    cnot q[4] | q[0];
                    cnot q[6] | q[1];
                    
                    cnot q[2] | q[0];
                    cnot q[5] | q[1];
                    cnot q[4] | q[3];
                    
                    cnot q[2] | q[1];
                    cnot q[6] | q[3];
                \end{yquant}
            \end{tikzpicture}}
            \caption{}
            \label{fig:steane_h_magic_circuit_1}
        \end{subfigure}
        \hfill
        \begin{subfigure}[b]{0.49\linewidth}
            \centering
            \resizebox{\linewidth}{!}{
            \begin{tikzpicture}
                \begin{yquant}
                    qubit {} q[7];
                    qubit {$\ket{+}$} q[+1];
                    qubit {$\ket{\makebox[\slength][c]{0}}$} q[+1];
                  
                    h q[6] | q[7];
                    cnot q[8] | q[7];
                    h q[5] | q[7];
                    h q[4] | q[7];
                    h q[3] | q[7];
                    h q[2] | q[7];
                    h q[1] | q[7];
                    cnot q[8] | q[7];
                    measure {$Z$} q[8];
                    h q[0] | q[7];
                    measure {$X$} q[7];
                \end{yquant}
            \end{tikzpicture}}
            \caption{}
            \label{fig:steane_h_magic_circuit_2}
        \end{subfigure}
        
        \vspace{0.5cm}
        
        \begin{subfigure}[b]{\linewidth}
            \centering
            \resizebox{\linewidth}{!}{
            \begin{tikzpicture}
                \begin{yquant}
                    qubit {} q[7];
                    qubit {$\ket{+}$} q[+1];
                    qubit {$\ket{\makebox[\slength][c]{0}}$} q[+2];
                  
                    cnot q[4] | q[7];
                    cnot q[8] | q[6];
                    cnot q[9] | q[5];
                    
                    cnot q[9] | q[7];
                    
                    cnot q[0] | q[7];
                    cnot q[8] | q[4];
                    cnot q[9] | q[1];
                    
                    cnot q[2] | q[7];
                    cnot q[8] | q[3];
                    cnot q[9] | q[6];
                    
                    cnot q[8] | q[7];
                    
                    cnot q[6] | q[7];
                    cnot q[8] | q[5];
                    cnot q[9] | q[2];

                    align q;
                    measure {$X$} q[7];
                    measure {$Z$} q[8];
                    measure {$Z$} q[9];

                    discard q[7-9];
                    init {$\ket{\makebox[\slength][c]{0}}$} q[7];
                    init {$\ket{+}$} q[8, 9];

                    cnot q[7] | q[4];
                    cnot q[6] | q[8];
                    cnot q[5] | q[9];
                    
                    cnot q[7] | q[9];
                    
                    cnot q[7] | q[0];
                    cnot q[4] | q[8];
                    cnot q[1] | q[9];
                    
                    cnot q[7] | q[2];
                    cnot q[3] | q[8];
                    cnot q[6] | q[9];
                    
                    cnot q[7] | q[8];
                    
                    cnot q[7] | q[6];
                    cnot q[5] | q[8];
                    cnot q[2] | q[9];

                    align q;
                    measure {$Z$} q[7];
                    measure {$X$} q[8];
                    measure {$X$} q[9];
                \end{yquant}
            \end{tikzpicture}}
            \caption{}
            \label{fig:steane_h_magic_circuit_3}
        \end{subfigure}
    \end{minipage}
    \hfill
    \begin{minipage}[b]{0.49\textwidth}
        \centering
        \begin{subfigure}[b]{\linewidth}
            \centering
            \includegraphics[width=1.0\linewidth]{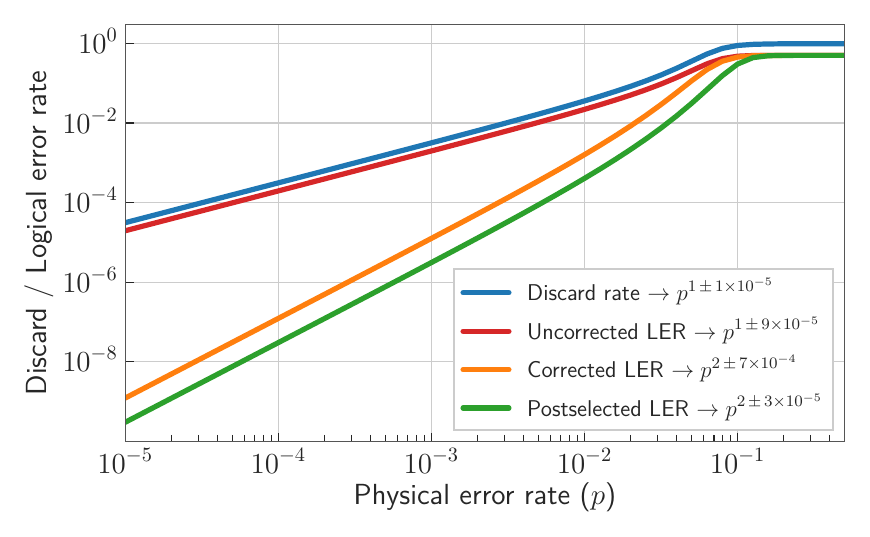}
            \caption{}
            \label{fig:plot_steane_h_magic}
        \end{subfigure}
    \end{minipage}

    \caption{Fault-tolerant (FT) magic state preparation of $\ket{\Bar{H}}$ in the Steane code, composed of the (a) non-FT magic state encoding, followed by the (b) FT measurement of the logical Hadamard operator and the (c) FT measurement of the six stabilisers. 
    (d) Combined (averaged) discard rate and logical error rates for the $Z$ and $X$ logical operators under the uncorrected (Eq.~\eqref{eq:uncorr_ler}), corrected (Eq.~\eqref{eq:corr_ler}) and postselected (Eq.~\eqref{eq:post_ler}) decoding strategies. The uncorrected $Z$ LER is $1.6 \, p^1 + 24.38 \, p^2 + O(p^3)$ and the postselected $Z$ LER is $2.15 \, p^2 + 80.46 \, p^3 + O(p^4)$. The uncorrected $X$ LER is $1.18 \, p^1 + 25.75 \, p^2 + O(p^3)$ and the postselected $X$ LER is $2.14 \, p^2 + 83.61 \, p^3 + O(p^4)$.}
    \label{fig:steane_h_magic}
\end{figure}

The circuit for the $\ket{\Bar{H}}$ magic state preparation in the Steane code is presented in Fig.~\ref{fig:steane_h_magic}. The complete circuit involves a non-FT encoding step, a measurement of the logical Hadamard operator $\Bar{H}$, and a syndrome extraction circuit~\cite{goto2016minimizing, chamberland2019fault}. In total, there are 15 physical qubits and 15 non-Clifford rotation gates: one for the original, unencoded $\ket{H}$ qubit, and two for implementing each of the 7 physical controlled-$H$ gates needed to measure $\Bar{H}$.

Fig.~\ref{fig:plot_steane_h_magic} shows LERs as small as $10^{-10}$ that are computed exactly, without Monte Carlo sampling errors or any other approximation. 
In the low physical error rate regime, the leading-order scaling of the LERs is clearly visible from a fit to the lowest three points, as shown in the legend of the figure, as well as the actual symbolic expressions provided in the caption. All LERs include postselection on the two ancillas in Fig.~\ref{fig:steane_h_magic_circuit_2}.

The postselected LER scales as $O(p^2)$ because the first step of the procedure is a non-FT encoding circuit, which can promote a single fault to a logical error, meaning that a second fault in the Hadamard measurement would not be able to detect such a logical error.
Crucially, this plot showcases the capability of our simulator to perform non-Clifford gates (15 rotations in this circuit) in conjunction with symbolic projections, leading to the computation of the circuit-level distance of a non-Clifford circuit.

\subsubsection{Decoding analysis} \label{sec:decoding_analysis}

\begin{table}[h]
    \centering
    \resizebox{\linewidth}{!}{%
    \begin{tabular}{|c|c|c|c|c|c|c|c|c|c|}
    \hline
    \multirow{2}{*}{Syndrome $\bm{s}$} &
      \multicolumn{3}{c|}{$p = 10^{-4}$} &
      \multicolumn{3}{c|}{$p = 10^{-3}$} &
      \multicolumn{3}{c|}{$p = 10^{-2}$} \\
    \cline{2-10}
    & Prob. & EV & Corr. LER & Prob. & EV & Corr. LER & Prob. & EV & Corr. LER \\
    \hline
    (0, 0, 0) & 0.99979 & $1\!-\!4\times\!10^{-12}$ & $2 \times 10^{-12}$ & 0.99785 &  $1\!-\!4\!\times\!10^{-9}$ & $2 \times 10^{-9}$ & 0.97744 &  $1\!-\!4\!\times\!10^{-6}$ & $2 \times 10^{-6}$ \\
    (0, 0, 1) & 0.00003 &  0.99979 & 0.00011 & 0.00027 &  0.99787 & 0.00107 & 0.00286 &  0.97890 & 0.01055 \\
    (0, 1, 0) & 0.00003 & -0.99899 & 0.00051 & 0.00027 & -0.98992 & 0.00504 & 0.00283 & -0.90405 & 0.04797 \\
    (0, 1, 1) & 0.00003 &  0.99925 & 0.00037 & 0.00027 &  0.99257 & 0.00372 & 0.00283 &  0.92870 & 0.03565 \\
    (1, 0, 0) & 0.00003 & -0.99947 & 0.00027 & 0.00027 & -0.99470 & 0.00265 & 0.00287 & -0.94959 & 0.02521 \\
    (1, 0, 1) & 0.00003 &  0.99845 & 0.00077 & 0.00027 &  0.98465 & 0.00767 & 0.00287 &  0.85640 & 0.07180 \\
    (1, 1, 0) & 0.00003 & -0.99872 & 0.00064 & 0.00027 & -0.98731 & 0.00635 & 0.00290 & -0.88190 & 0.05905 \\
    (1, 1, 1) & 0.00005 &  0.99992 & 0.00004 & 0.00053 &  0.99919 & 0.00040 & 0.00541 &  0.99149 & 0.00425 \\
    \hline
    \multirow{1}{*}{Corrected LER} &
      \multicolumn{3}{c|}{$7.32 \times 10^{-8}$} &
      \multicolumn{3}{c|}{$7.33 \times 10^{-6}$} &
      \multicolumn{3}{c|}{$7.41 \times 10^{-4}$} \\
    \hline
    \end{tabular}
    }
    \caption{Circuit-level maximum-likelihood look-up table (CL-ML-LUT) information for the $\ket{\Bar{0}}$ state preparation of the Steane code for three physical error rates $p$. On the accepted state $\rho$ after the state preparation we noiselessly project onto the $\bm{s}$ syndrome of the three $Z$-type stabilisers and apply the destabilisers for $\bm{s}$, obtaining a state $\mathsf{D}_{\bm{s}}(\rho)$ in the code space. For the bitstring of each syndrome, 0 represents a $+1$ stabiliser measurement, and 1 represents a $-1$ measurement.
    From left to right: syndrome $\bm{s}$, probability of measuring $\bm{s}$, expectation value of the logical operator $\Bar{Z}$, and corrected logical error rate (Eq.~\eqref{eq:corr_ler}).
    Positive or negative expectation values indicate that the optimal correction programs apply an identity or a logical $\Bar{X}$ correction.
    The last row indicates the average corrected LER over all syndromes.
    In the noiseless case, the probability of the trivial syndrome $\bm{s}=+1$ is 1, the expectation value is 1, and the LERs are 0.}
    \label{tab:steane_8}
\end{table}

The information presented in Table~\ref{tab:steane_8} allows us to highlight syndromes such as \texttt{(1, 0, 1)}, that have larger LER after CL-ML correction than the others, making them more unreliable to decode.
These syndromes are the strongest candidates to be discarded instead of attempting to correct upon their observation.
In this case, the bitstrings are used to represent the fact that \texttt{0} is a projection on the $+1$ eigenstate and \texttt{1} is a projection on the $-1$ eigenstate.

\begin{figure}[h!]
    \centering
    \begin{subfigure}[c]{0.49\textwidth}
        \centering
        \includegraphics[width=1.0\linewidth]{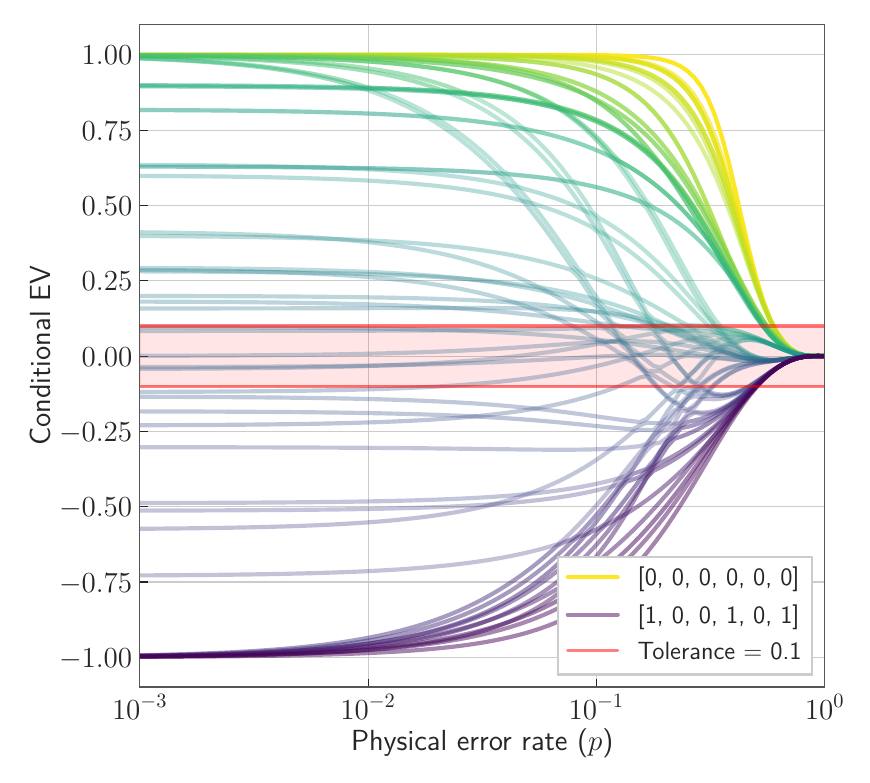}
        \caption{}
        \label{fig:syndrome_plot_smooth}
    \end{subfigure}
    \hfill
    \begin{subfigure}[c]{0.49\textwidth}
        \centering
        \includegraphics[width=1.0\linewidth]{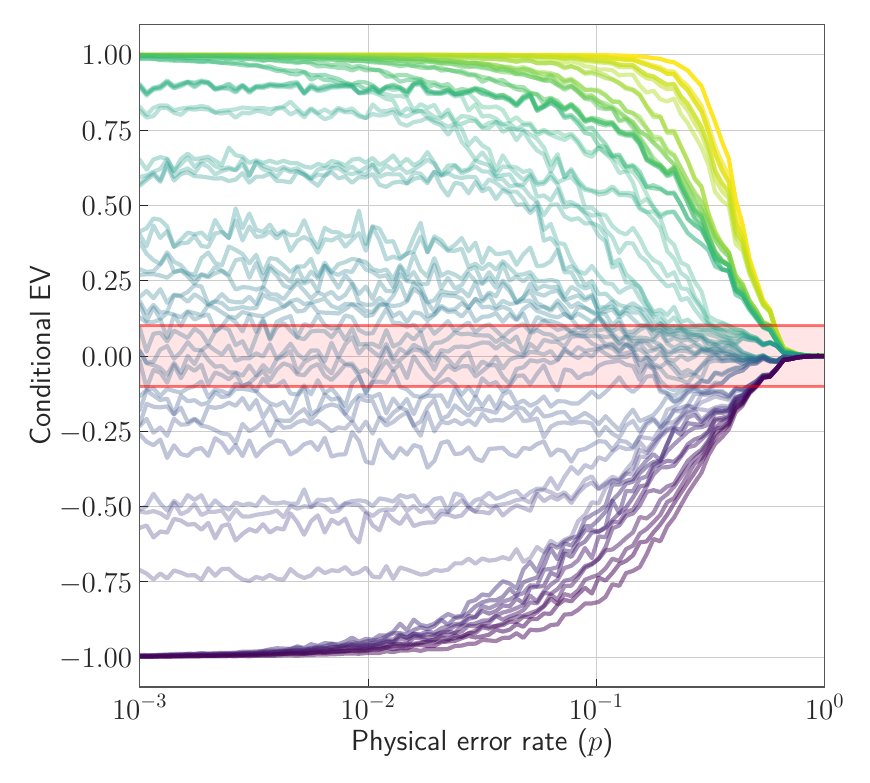}
        \caption{}
        \label{fig:syndrome_plot_fluctuating}
    \end{subfigure}
    \caption{Expectation values (EVs) of the logical operator $\Bar{Z}$ of the prepared state $\ket{\Bar{0}}$ in the Steane code as a function of the physical error rate $p$ for every syndrome $\bm{s}$.
    On the accepted state $\rho$ after the state preparation, we noiselessly project onto the $\bm{s}$ syndrome of all the six stabilisers and apply the destabilisers for $\bm{s}$, obtaining a state $\mathsf{D}_{\bm{s}}(\rho)$ in the code space.
    Positive or negative expectation values indicate that the optimal correction programs apply an identity or a logical $\Bar{X}$ correction.
    The legend shows the top syndrome, with the highest EV, and the bottom syndrome, with the lowest EV. Syndromes with higher probabilities have more solid lines, and syndromes with lower probabilities have more transparent lines. (a) The conditional EVs are calculated exactly at the corresponding physical error rate values on the $x$-axis. (b) The conditional EVs are calculated at physical error rates sampled from a distribution centred around the corresponding values on the $x$-axis, emulating hardware miscalibration.
    The horizontal red lines delimit the red, shaded region where the syndromes are most unreliable to decode, with expectation values close to 0. The postselected $Z$ EV (the top, all-0 syndrome) is $1 - 0.43 \, p^3 - 12.88 \, p^4 + O(p^5)$.}
    \label{fig:syndrome_plot}
\end{figure}

The information presented in Fig.~\ref{fig:syndrome_plot} can be used to decode better in practice.
Note that in Fig.~\ref{fig:syndrome_plot_smooth} some expectation values change sign for high values of the physical error rate $p$, and recall that the sign determines the optimal correction.
Therefore, this figure shows the importance of designing any decoding strategy that takes circuit-level and noise information into account at the physical error rate that best describes the hardware.
Unfortunately, there is always some uncertainty in the noise characterisation of the hardware that can lead to incorrectly designing the decoder, making the correction of some syndromes unreliable.
We indicate by the red band one illustrative example of tolerance for this miscalibration: if the expectation value for a syndrome falls inside that band, its sign is unreliable and should therefore be discarded instead of corrected.
To further stress the importance of the miscalibration we insert random fluctuations of the physical error rate in Fig.~\ref{fig:syndrome_plot_fluctuating}, showing that some syndromes might change sign (changing the optimal correction) inside the tolerance band.
We find that, at $p = 10^{-3}$, eliminating the syndromes within the red tolerance region, which correspond to EVs in the $[-0.1, 0.1]$ range and have a cumulative probability of less than one in a million, leads to the LER reducing by roughly 10\%. If we remove all the syndromes whose EVs do not converge to $\pm1$, so everything but those at the very top and at the very bottom of the figure, having a cumulative probability of less than 1 in 100,000, then the LER halves. The cumulative probability of the removed syndromes contributes to the total discard rate, so it is preferable for it to be as small as possible.
This example showcases the analysis power of SyQMA, obtaining exact expressions for all the syndromes and helping us decide which syndromes are most affected by variations in noise and best to be discarded in order to improve the corrected LER.

\begin{table}[h]
    \centering
    \resizebox{\linewidth}{!}{%
    \begin{tabular}{|c|c|c|c|c|c|c|}
    \hline
    \rule{0pt}{3ex} Syndrome $\bm{s}$ & Prob. & EV $\Bar{Z}$ & EV $\Bar{X}$ & Correction & Corr. LER $\Bar{Z}$ & Corr. LER $\Bar{X}$ \\
    \hline
    (0, 0, 0, 0, 0, 0) & 0.99682609 &  0.707102 &  0.707102 & $\Bar{I}$ & 0.000002 & 0.000002 \\
    (0, 1, 1, 0, 0, 0) & 0.00053503 & -0.703658 &  0.707056 & $\Bar{X}$ & 0.001724 & 0.000025 \\
    (1, 0, 1, 0, 0, 0) & 0.00026722 & -0.705806 &  0.707075 & $\Bar{X}$ & 0.000650 & 0.000016 \\
    (0, 0, 0, 1, 1, 1) & 0.00026700 &  0.707099 & -0.706928 & $\Bar{Z}$ & 0.000004 & 0.000090 \\
    (1, 1, 0, 0, 0, 0) & 0.00020049 & -0.704998 &  0.707044 & $\Bar{X}$ & 0.001055 & 0.000031 \\
    $\cdots$ & $\cdots$ & $\cdots$ & $\cdots$ & $\cdots$ & $\cdots$ & $\cdots$ \\
    (0, 0, 1, 1, 0, 0) & 0.00000091 & -0.215412 &  0.073402 & $\Bar{X}$ & 0.245847 & 0.316852 \\
    $\cdots$ & $\cdots$ & $\cdots$ & $\cdots$ & $\cdots$ & $\cdots$ & $\cdots$ \\
    (1, 1, 0, 0, 1, 1) & 0.00000010 & -0.189846 & -0.445545 & $\Bar{Y}$ & 0.258630 & 0.130781 \\
    (1, 1, 1, 0, 1, 1) & 0.00000007 & -0.266687 &  0.163829 & $\Bar{X}$ & 0.220210 & 0.271639 \\
    (1, 0, 0, 0, 1, 1) & 0.00000007 & -0.344049 &  0.165668 & $\Bar{X}$ & 0.181529 & 0.270719 \\
    (0, 0, 1, 0, 1, 1) & 0.00000007 &  0.135838 & -0.419393 & $\Bar{Z}$ & 0.285634 & 0.143857 \\
    (0, 1, 0, 0, 1, 1) & 0.00000007 & -0.230370 & -0.420263 & $\Bar{Y}$ & 0.238368 & 0.143422 \\
    \hline
    \multirow{1}{*}{Acceptance Rate} &
      \multicolumn{6}{c|}{0.924047} \\
    \hline
    \multirow{1}{*}{Uncorrected LER} &
      \multicolumn{6}{c|}{0.001406} \\
    \hline
    \multirow{1}{*}{Corrected LER} &
      \multicolumn{6}{c|}{0.000009} \\
    \hline
    \multirow{1}{*}{Postselected LER} &
      \multicolumn{6}{c|}{0.000002} \\
    \hline
    \end{tabular}
    }
    \caption{Circuit-level maximum-likelihood look-up table (CL-ML-LUT) for the magic $\ket{\Bar{H}}$ state preparation of the Steane code for the physical error rate $p=10^{-3}$. On the accepted state $\rho$ after the state preparation, we noiselessly project onto the $\bm{s}$ syndrome of the six stabilisers and apply the destabilisers for $\bm{s}$, obtaining a state $\mathsf{D}_{\bm{s}}(\rho)$ in the code space.
    From left to right: syndrome $\bm{s}$, probability of measuring $\bm{s}$, expectation value of the logical operators $\Bar{Z}$ and $\Bar{X}$, optimal logical correction, and corrected logical error rates (Eq.~\eqref{eq:corr_lerO}).
    The LERs of $\Bar{Z}$ ($\Bar{X}$) are the probabilities that $\mathsf{D}_{\bm{s}}(\rho)$ carries logical $\Bar{X}$ or $\Bar{Y}$ ($\Bar{Z}$ or $\Bar{Y}$) error channels.
    The correction is given by the signs of the expectation values.
    The last four rows indicate, from top to bottom, the acceptance rate $\Tr(\rho)$, the uncorrected (Eq.~\eqref{eq:uncorr_ler}), corrected (Eq.~\eqref{eq:corr_ler}) and postselected (Eq.~\eqref{eq:post_ler}) LERs, averaged over all syndromes.
    In the noiseless case, the probability of the trivial syndrome $\bm{s}=+1$ is 1, the expectation values are $1/\sqrt{2}$, and the LERs are 0.}
    \label{tab:steane_magic}
\end{table}

Table~\ref{tab:steane_magic} shows the CL-ML-LUT data for the FT preparation of the $\ket{\Bar{H}}$ magic state on the Steane code. 
The trivial syndrome is by far the most probable, with the lowest $Z$ and $X$ LERs. 
The second most probable syndrome \texttt{(0, 1, 1, 0, 0, 0)} has only $Z$-type stabilisers excited, which might indicate a larger presence of $X$-type errors than of $Z$- and $Y$-type.
Correspondingly, the LER for $\Bar{Z}$ is almost two orders of magnitude larger than for $\Bar{X}$.
Roughly halfway across the list of syndromes ordered by probability, all syndromes from \texttt{(0, 0, 1, 1, 0, 0)} onwards exhibit high LERs.
This makes them the best candidates to be discarded instead of corrected.
The provided LUTs could be directly used for decoding in a real hardware experiment if the noise model we employ were similar to the one on the real hardware.

\subsection{$\llbracket15,1,3\rrbracket$ 3D colour code} \label{sec:15_1_3_clifford}
This section presents our results for the $\llbracket15,1,3\rrbracket$ Reed-M\"uller 3D colour code.
Unlike the Steane code, this code is not self-dual, but it features a logical non-Clifford $\Bar{T}=\exp(-i\pi \Bar{Z}/8)$ implemented transversally via single-qubit gates.
We present it in our results to demonstrate the simulation of stabiliser and non-stabiliser states at the limit size of density matrix simulators, that, like SyQMA, can perform exact simulation, but unlike SyQMA, suffer from exponential memory consumption. 
The code has 10~$Z$-type and 4~$X$-type stabiliser generators, resulting in asymmetric distances $d_Z = 5$ and $d_X = 3$ for the respective logical operators.

\subsubsection{Fault-tolerant Clifford state preparation}

\begin{figure}[h]
    \centering
    \begin{subfigure}[c]{0.445\textwidth}
        \centering
        \resizebox{\linewidth}{!}{
        \begin{tikzpicture}
            \begin{yquant}
                qubit {$\ket{+}$} q;
              qubit {$\ket{\makebox[\slength][c]{0}}$} q[+7];
              qubit {$\ket{+}$} q[+1];
              qubit {$\ket{\makebox[\slength][c]{0}}$} q[+1];
              qubit {$\ket{+}$} q[+1];
              qubit {$\ket{\makebox[\slength][c]{0}}$} q[+3];
              qubit {$\ket{+}$} q[+1];
              
              cnot q[7] | q[14];
              cnot q[3] | q[10];
              cnot q[1] | q[8];
              cnot q[12] | q[14];
              cnot q[9] | q[10];
              cnot q[5] | q[7];
              cnot q[2] | q[3];
              cnot q[1] | q[0];
              cnot q[13] | q[14];
              cnot q[11] | q[12];
              cnot q[9] | q[8];
              cnot q[6] | q[7];
              cnot q[4] | q[5];
              cnot q[3] | q[0];
              cnot q[2] | q[1];
              cnot q[13] | q[10];
              cnot q[12] | q[8];
              cnot q[7] | q[0];
              cnot q[11] | q[9];
              cnot q[6] | q[3];
              cnot q[4] | q[1];
              cnot q[5] | q[2];
        
              [after=q]
              qubit {$\ket{+}$} verification_flag[1];
              cnot verification_flag[0] | q[0];
              cnot verification_flag[0] | q[5];
              cnot verification_flag[0] | q[11];
              measure verification_flag[0];
            \end{yquant}
        \end{tikzpicture}}
        \caption{}
        \label{fig:circuit_15_1_3}
    \end{subfigure}
    \hfill
    \begin{subfigure}[c]{0.545\textwidth}
        \centering
        \includegraphics[width=1.0\linewidth]{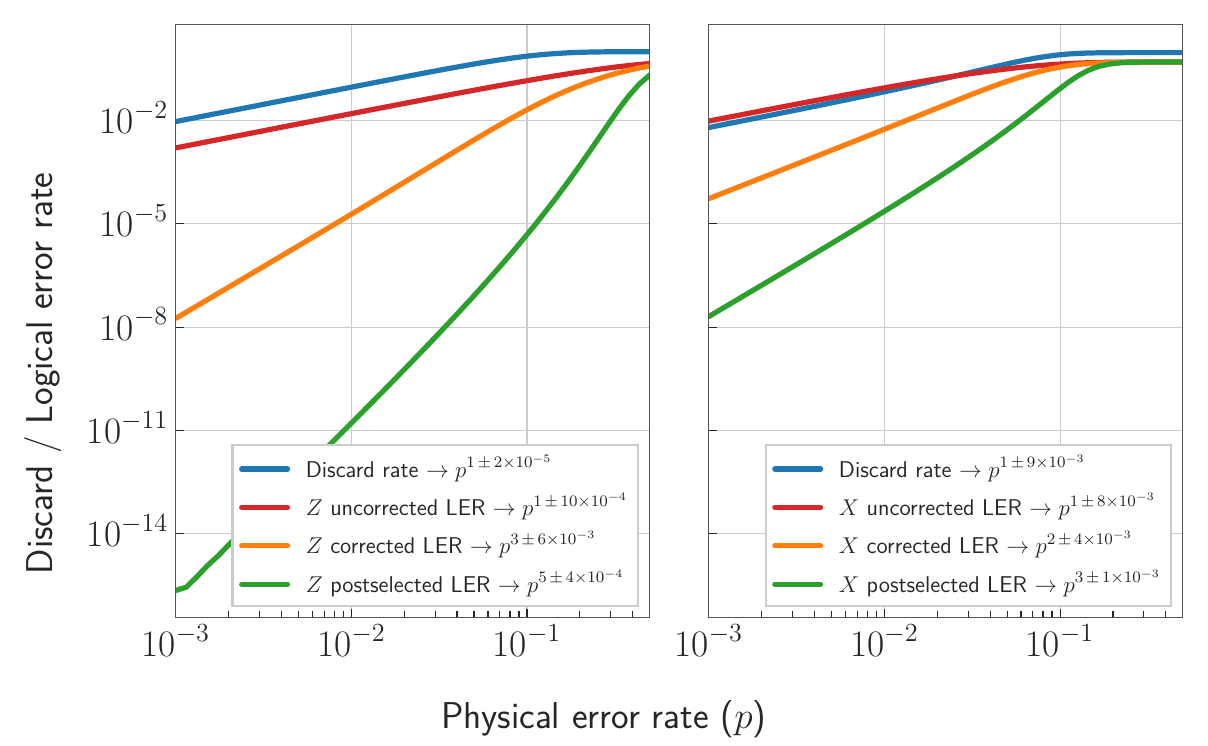}
        \caption{}
        \label{fig:15_1_3_ler_zx}
    \end{subfigure}
    \caption{Fault-tolerant preparation of Clifford states in the $\llbracket15,1,3\rrbracket$ 3D colour code. (a) Verified preparation gadget for $\ket{\Bar{0}}$, where the accepted branch postselects on the flag outcome $+1$. The $\ket{\Bar{+}}$ gadget used in panel~(b) is obtained from Fig.~\ref{fig:circuit_15_1_3_T} by removing the final transversal $T^\dagger$ layer. (b) Exact discard rates and logical error rates (LERs) for $\Bar{Z}$ on $\ket{\Bar{0}}$ (left panel) and $\Bar{X}$ on $\ket{\Bar{+}}$ (right panel) under the uncorrected (Eq.~\eqref{eq:uncorr_ler}), corrected (Eq.~\eqref{eq:corr_ler}), and postselected (Eq.~\eqref{eq:post_ler}) decoding strategies. The uncorrected $Z$ LER is $1.6 \, p^1 + 10.24 \, p^2 + O(p^3)$ and the postselected $Z$ LER is $0.14 \, p^5 + 7.08 \, p^6 + O(p^7)$. The uncorrected $X$ LER is $9.73 \, p^1 - 20.87 \, p^2 + O(p^3)$ and the postselected $X$ LER is $19.95 \, p^3 + 673.46 \, p^4 + O(p^5)$.}
    \label{fig:15_1_3_clifford}
\end{figure}

The circuit for the $\ket{\Bar{0}}$ state preparation and syndrome measurements in the $\llbracket15,1,3\rrbracket$ code is presented in Fig.~\ref{fig:circuit_15_1_3}, while the circuit for the $\ket{\Bar{+}}$ state preparation and syndrome measurements is the one in Fig.~\ref{fig:circuit_15_1_3_T} \cite{peham2025automated} without the final layer of non-Clifford gates. Fig.~\ref{fig:15_1_3_ler_zx} shows the exact calculation of LERs down to $10^{-15}$ for the $\ket{\Bar{0}}$ and $\ket{\Bar{+}}$ FT state preparations of the $\llbracket15,1,3\rrbracket$ code.

In the low error rate regime, we recover the asymptotic scaling of the LER with the leading order in $p$, which differs between the logical Pauli operators due to the asymmetric stabiliser structure: $d_Z = 5$ and $d_X = 3$. Specifically, for the $\Bar{Z}$ operator, the leading orders are $O(p)$ for the uncorrected LER, $O(p^3)$ for the corrected LER after CL-ML decoding, and $O(p^5)$ for the fully postselected LER. For the $\Bar{X}$ operator, the leading orders are $O(p)$ for the uncorrected LER, $O(p^2)$ for the corrected LER after CL-ML decoding, and $O(p^3)$ for the fully postselected LER. Since the $Z$-distance is higher than the $X$-distance, with more $Z$ stabilisers (10) than $X$ stabilisers (4), the LER is much lower on the left panel than on the right panel. All values shown are calculated exactly, with no shot noise. The actual symbolic expressions are provided in the caption.

\subsubsection{Fault-tolerant magic state preparation}

\begin{figure}[h]
    \centering
    \begin{subfigure}[c]{0.49\textwidth}
        \centering
        \resizebox{\linewidth}{!}{
        \begin{tikzpicture}
            \begin{yquant}
                qubit {$\ket{\makebox[\slength][c]{0}}$} q[2];
                qubit {$\ket{+}$} q[+1];
                qubit {$\ket{\makebox[\slength][c]{0}}$} q[+1];
                qubit {$\ket{+}$} q[+1];
                qubit {$\ket{\makebox[\slength][c]{0}}$} q[+1];
                qubit {$\ket{+}$} q[+1];
                qubit {$\ket{\makebox[\slength][c]{0}}$} q[+3];
                qubit {$\ket{+}$} q[+1];
                qubit {$\ket{\makebox[\slength][c]{0}}$} q[+2];
                qubit {$\ket{+}$} q[+1];
                qubit {$\ket{\makebox[\slength][c]{0}}$} q[+1];
        
                cnot q[11]|q[13];
                cnot q[12]|q[11];
                cnot q[5]|q[6];
                cnot q[3]|q[10];
                cnot q[1]|q[4];
                cnot q[9]|q[11];
                cnot q[8]|q[12];
                cnot q[7]|q[4];
                cnot q[5]|q[3];
                cnot q[1]|q[2];
                cnot q[14]|q[7];
                cnot q[13]|q[6];
                cnot q[9]|q[2];
                cnot q[8]|q[1];
                cnot q[0]|q[3];
                cnot q[4]|q[5];
                cnot q[14]|q[13];
                cnot q[10]|q[9];
                cnot q[0]|q[1];
                cnot q[12]|q[4];
                cnot q[7]|q[6];
                cnot q[11]|q[5];
                cnot q[3]|q[2];
                
                [after=q]
                qubit {$\ket{+}$} x_anc[1];
                cnot q[1]|x_anc[0];
                cnot q[3]|x_anc[0];
                cnot q[5]|x_anc[0];
                cnot q[7]|x_anc[0];
                cnot q[9]|x_anc[0];
                cnot q[12]|x_anc[0];
                cnot q[13]|x_anc[0];
                
                measure {$X$} x_anc[0];
                
                [after=q]
                qubit  {$\ket{\makebox[\slength][c]{0}}$} z_anc[2];
                
                cnot z_anc[0]|q[0];
                
                [after=z_anc]
                qubit {$\ket{+}$} a15[1];
                cnot z_anc[0]|a15[0];
                cnot z_anc[0]|q[5];
                cnot z_anc[0]|q[9];
                cnot z_anc[0]|a15[0];
                cnot z_anc[0]|q[14];
                measure {$Z$} z_anc[0];
                cnot z_anc[1]|q[3];
                [after=z_anc]
                qubit {$\ket{+}$} a16[1];
                cnot z_anc[1]|a16[0];
                cnot z_anc[1]|q[7];
                cnot z_anc[1]|q[9];
                cnot z_anc[1]|a16[0];
                cnot z_anc[1]|q[12];
                
                measure {$X$} a16[0];
                measure {$X$} a15[0];
                measure {$Z$} z_anc[1];

                hspace {2mm} q[0-14];
                
                [this subcircuit box style={rounded corners, fill=blue!20, label=below:\textsc{$T^\dagger$-Layer}}]
                subcircuit {
                    % Inside the subcircuit, you must declare the internal input wires.
                    % Since we are passing 15 wires in, we declare a vector of 15 qubits.
                    qubit {} inner_q[15];
                    
                    % Apply the T-dagger to each individual internal register.
                    % (Omitting parentheses around the index applies it to each wire individually)
                    box {$T^\dagger$} inner_q[0-14];
                    
                } (q[0-14]); % Pass the outer registers q[0] through q[14] into the block
            \end{yquant}
        \end{tikzpicture}}
        \caption{}
        \label{fig:circuit_15_1_3_T}
    \end{subfigure}
    \hfill
    \begin{subfigure}[c]{0.49\textwidth}
        \centering
        \includegraphics[width=1.0\linewidth]{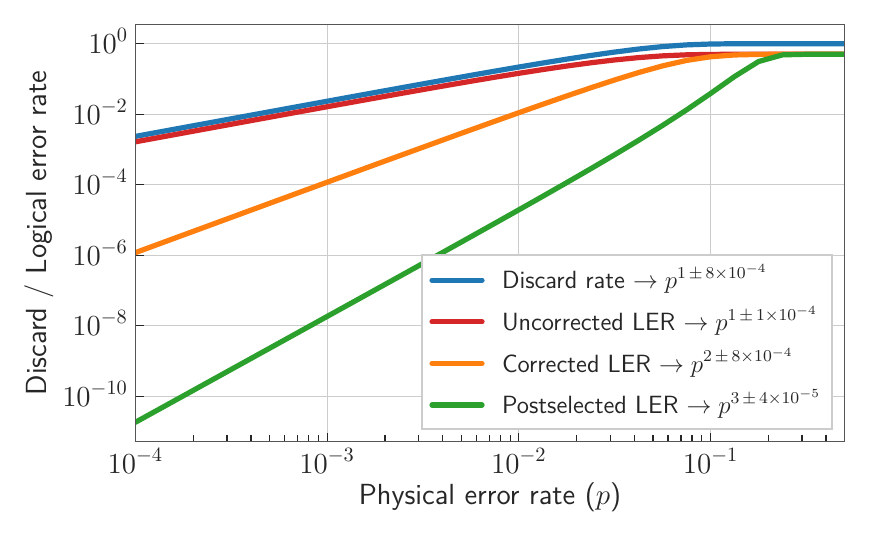}
        \caption{}
        \label{fig:15_1_3_T_ler_x}
    \end{subfigure}
    \caption{Fault-tolerant magic-state preparation of $\ket{\Bar{T}}$ in the $\llbracket15,1,3\rrbracket$ 3D colour code. (a) The verified $\ket{\Bar{+}}$ preparation and syndrome-extraction gadget, followed by the transversal $T^\dagger$ layer on all 15 data qubits. (b) Exact combined (averaged) discard rate and logical error rates for the $\Bar{X}$ and $\Bar{Y}$ operators under the uncorrected (Eq.~\eqref{eq:uncorr_ler}), corrected (Eq.~\eqref{eq:corr_ler}), and postselected (Eq.~\eqref{eq:post_ler}) decoding strategies.}
    \label{fig:15_1_3_magic}
\end{figure}

The circuit for the $\ket{\Bar{T}}$ FT state preparation in the $\llbracket15,1,3\rrbracket$ code is presented in Fig.~\ref{fig:circuit_15_1_3_T}. The procedure starts with the $\ket{\Bar{+}}$ FT state preparation circuit, taken from \cite{peham2025automated} and already presented in Section~\ref{sec:15_1_3_clifford}. The code admits a transversal $T$ gate, which is then applied to the $\ket{\Bar{+}}$ state by performing a physical $T^\dagger$ gate on every data qubit. In total, there are 20 physical qubits and 15 non-Clifford rotation gates. Unlike the Steane code magic state preparation, this procedure is fully FT: the non-Clifford layer is transversal and therefore does not introduce correlated faults.

Fig.~\ref{fig:15_1_3_T_ler_x} shows the exact calculation of combined (averaged) LERs as low as $10^{-14}$ for the $\Bar{X}$ and $\Bar{Y}$ operators for the FT state preparation of the $\ket{\Bar{T}}$ magic state in the $\llbracket15,1,3\rrbracket$ 3D colour code. The discard rate scales as $O(p)$, representing the probability of not measuring the ancillas and the stabilisers in the $+1$ eigenstates. In the low physical error regime, the expected asymptotic scaling for a $d = 3$ code is recovered, with leading terms of $O(p)$ for the uncorrected LER, $O(p^2)$ for the corrected LER after CL-ML decoding, and $O(p^3)$ for the fully postselected LER. The full preservation of the code distance under postselection, in contrast to the $O(p^2)$ postselected scaling observed for the partially-FT Steane code magic state, reflects the transversal nature of the $T$ gate on this code.

\subsubsection{Memory and time overhead} \label{sec:memory_time}

\begin{figure}[h]
    \centering
    \includegraphics[width=0.8\linewidth]{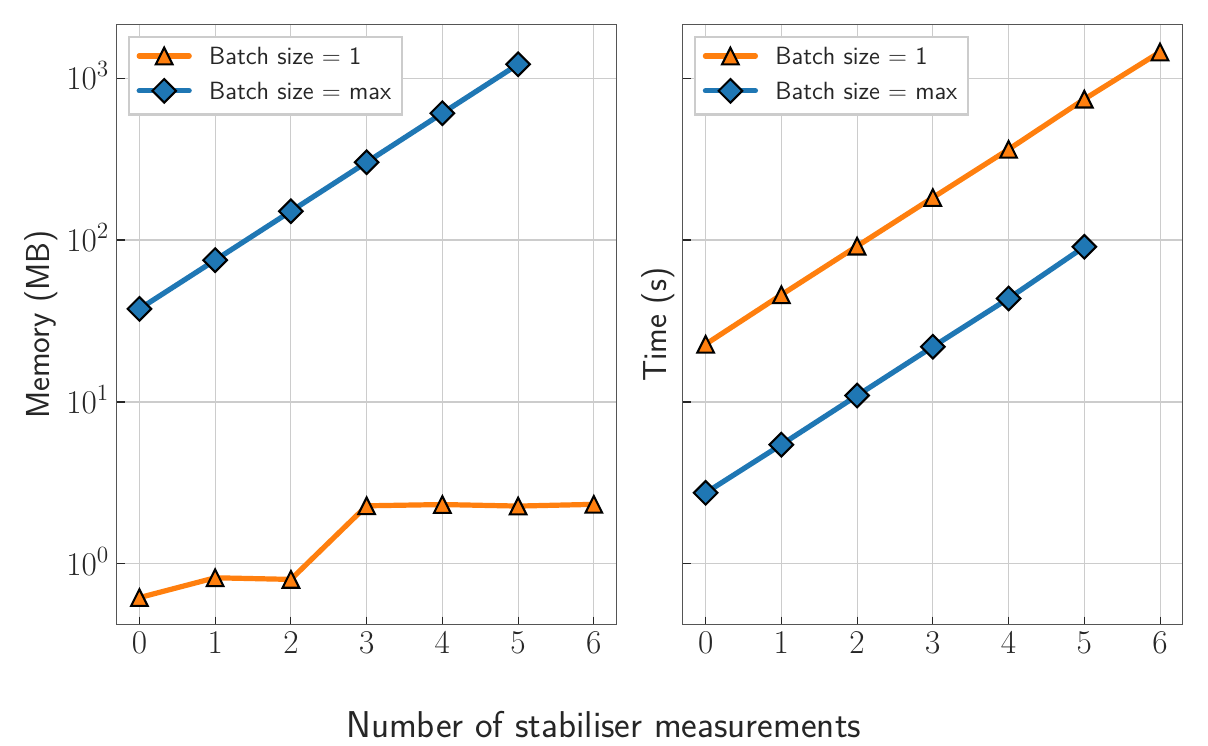}
    \caption{Comparison of memory and runtime between the computation mode with the largest batch size and the mode with minimal batch size of one for calculating the $\Bar{X}$ operator in the $\ket{\Bar{T}}$ magic state of the $\llbracket15,1,3\rrbracket$ colour code.}
    \label{fig:15_1_3_T_memory_time}
\end{figure}

Fig.~\ref{fig:15_1_3_T_memory_time} shows the total time and the peak memory taken for the calculation of the symbolic EV for the $\Bar{X}$ operator for the $\ket{\Bar{T}}$ magic state in the $\llbracket15,1,3\rrbracket$ 3D colour code with an increasing number of stabiliser measurements, each doubling the number of terms in the expression for the EV. The base case, with no stabiliser measurements, is exactly the circuit shown in Fig.~\ref{fig:circuit_15_1_3_T}.

The plots highlight the two regimes in which SyQMA can be operated: the orange line (batch size of 1) represents the regime where each of the exponentially-many terms in the EV can be generated sequentially with only a polynomial amount of memory, while the blue line (maximum batch size) shows the regime where all the exponentially-many terms are generated at once, scaling similarly to the memory in a full density matrix simulation. By being able to generate the terms of the final EV individually (batch size of 1), we get a huge reduction in the memory used, for only a modest increase in the computational time taken (an order of magnitude in this case). Once again, we remark SyQMA's ability to calculate exact, noisy EVs with only polynomial memory and with no exponential dependence on the total number of qubits.

\subsection{$\llbracket17,1,5\rrbracket$ colour code}

The $\llbracket17,1,5\rrbracket$ code is a 4.8.8 2D colour code~\cite{bombin2006colorcode} that encodes one logical qubit in 17 physical qubits at distance $d=5$, making it the highest-distance code analysed in this work.
With $t=(d-1)/2=2$, it can correct any two faults, so a FT state preparation must ensure that logical errors require at least $t+1=3$ faults.

\begin{figure}[h!]
    \centering
    \begin{subfigure}[c]{\linewidth}
        \centering
        \resizebox{\linewidth}{!}{
        \begin{tikzpicture}
            \begin{yquant}[operator/separation=2mm]
                qubit {$q_{\idx}: \ket{0}$} q[38];
    
                h q[3, 4, 5, 6, 7, 8, 14, 17, 19, 21, 22, 23, 25, 26, 28, 30, 37];
                
                align q;
    
                cnot q[2] | q[21];
                cnot q[1] | q[21];
                cnot q[0] | q[21];
                cnot q[36] | q[8];
                cnot q[15] | q[25];
                cnot q[36] | q[21];
                cnot q[16] | q[26];
                cnot q[33] | q[5];
                cnot q[33] | q[21];
                cnot q[29] | q[25];
                cnot q[2] | q[21];
                cnot q[29] | q[26];
                cnot q[34] | q[6];
                cnot q[34] | q[25];
                cnot q[34] | q[1];
                cnot q[24] | q[14];
                cnot q[33] | q[26];
                cnot q[32] | q[4];
                cnot q[35] | q[21];
                cnot q[33] | q[25];
                cnot q[32] | q[1];
                cnot q[35] | q[7];
                cnot q[35] | q[26];
                cnot q[27] | q[17];
                cnot q[24] | q[21];
                cnot q[15] | q[25];
                cnot q[1] | q[21];
                cnot q[16] | q[26];
                cnot q[9] | q[37];
                cnot q[27] | q[21];
                cnot q[0] | q[21];
                cnot q[36] | q[37];
                cnot q[31] | q[37];
                cnot q[18] | q[28];
                cnot q[10] | q[22];
                cnot q[32] | q[37];
                cnot q[31] | q[28];
                cnot q[32] | q[22];
                cnot q[9] | q[37];
                cnot q[31] | q[3];
                cnot q[29] | q[19];
                cnot q[32] | q[28];
                cnot q[24] | q[22];
                cnot q[27] | q[28];
                cnot q[34] | q[6];
                cnot q[18] | q[28];
                cnot q[31] | q[22];
                cnot q[20] | q[30];
                cnot q[32] | q[4];
                cnot q[10] | q[22];
                cnot q[34] | q[30];
                cnot q[29] | q[30];
                cnot q[35] | q[12];
                cnot q[35] | q[30];
                cnot q[24] | q[12];
                cnot q[20] | q[30];
                cnot q[33] | q[5];
                cnot q[36] | q[8];
                cnot q[13] | q[23];
                cnot q[12] | q[23];
                cnot q[11] | q[23];
                cnot q[31] | q[3];
                cnot q[35] | q[7];
                cnot q[24] | q[14];
                cnot q[29] | q[19];
                cnot q[31] | q[23];
                cnot q[34] | q[23];
                cnot q[27] | q[17];
                cnot q[13] | q[23];
                cnot q[36] | q[23];
                cnot q[27] | q[23];
                cnot q[12] | q[23];
                cnot q[33] | q[23];
                cnot q[11] | q[23];
    
                align q;

                h q[3-8, 14, 17, 19];
                measure q[0-19];
            \end{yquant}
        \end{tikzpicture}}
        \caption{}
        \label{fig:circuit_17_1_5}
    \end{subfigure}
    
    \begin{subfigure}[c]{\linewidth}
        \centering
        \includegraphics[width=0.59\linewidth]{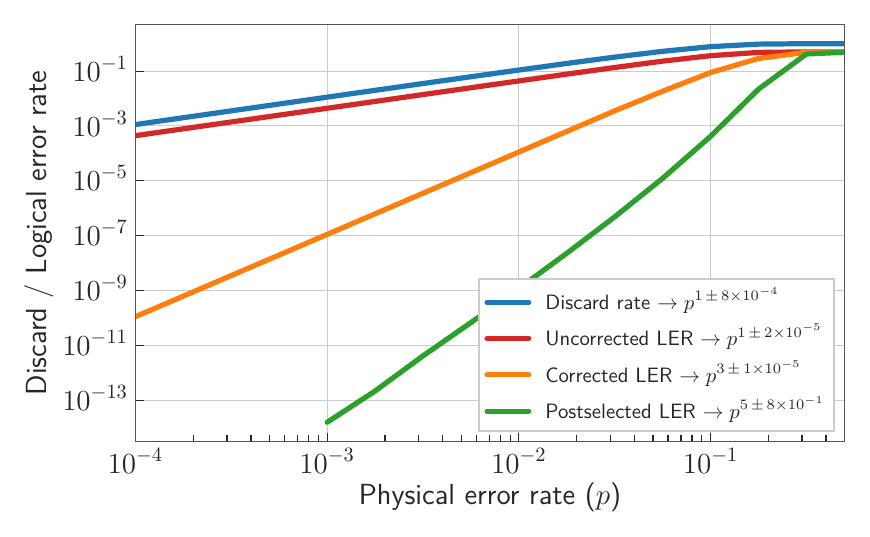}
        \caption{}
        \label{fig:17_1_5_ler_z}
    \end{subfigure}
    \caption{Fault-tolerant preparation of $\ket{\Bar{0}}$ in the $\llbracket17,1,5\rrbracket$ 2D colour code. (a) Flagged preparation and syndrome extraction circuit adapted from~\cite{forlivesi2025flag}, where the accepted branch postselects on all measured ancilla and syndrome outcomes being $+1$. (b) Exact discard rate and $\Bar{Z}$ logical error rates under the uncorrected (Eq.~\eqref{eq:uncorr_ler}), corrected (Eq.~\eqref{eq:corr_ler}), and postselected (Eq.~\eqref{eq:post_ler}) decoding strategies.}
    \label{fig:17_1_5_clifford}
\end{figure}

\subsubsection{Fault-tolerant Clifford state preparation}

The circuit for the FT preparation of $\ket{\Bar{0}}$ and the accompanying syndrome measurements in the $\llbracket17,1,5\rrbracket$ code is shown in Fig.~\ref{fig:circuit_17_1_5}, following Ref.~\cite{forlivesi2025flag}. The gadget uses 38 physical qubits in total, with 17 data qubits and the remainder used for flags and syndrome extraction. The circuit is entirely Clifford, so the exponential cost arises solely from the deterministic measurements that become non-deterministic under noise.

Fig.~\ref{fig:17_1_5_ler_z} shows the exact calculation of LERs as low as $10^{-14}$ for the $\Bar{Z}$ operator for the FT state preparation of $\ket{\Bar{0}}$ in the $\llbracket17,1,5\rrbracket$ code. The discard rate scales as $O(p)$, consistent with the FT property that single faults are detected and discarded. In the low physical error regime, the expected asymptotic scaling for a $d = 5$ code is recovered, with leading terms of $O(p)$ for the uncorrected LER, $O(p^3)$ for the corrected LER after CL-ML decoding, and $O(p^5)$ for the fully postselected LER. Notably, the postselected LER for the first two physical error rates falls below typical double-precision floating-point limits near unity ($\sim\!10^{-15}$), demonstrating that SyQMA is limited only by hardware numerical precision rather than by sampling noise or other approximation errors.

\section{Outlook}
\label{sec:outlook}
In this work, we have presented SyQMA, a universal, exact, symbolic, and memory-efficient classical simulator particularly suited for quantum error correction (QEC) tasks. 
We demonstrated simulation, decoding, computation of LERs, and verification of fault-distance preservation for fault-tolerant (FT) stabiliser and magic state preparation across several small QEC codes.

Future work includes expanding SyQMA to make it a more complete simulator. 
For instance, the set of quantum operations could be extended to more general noise sources, such as amplitude damping and leakage. 
We could also compute state overlaps, channel distances, or entanglement properties.
For QEC, the decoding framework may be extended beyond state preparation, e.g., by computing channel fidelities between noiseless and noisy versions of QEC programs like QEC cycles or lattice surgery. 
Decoding could also incorporate correction programs beyond Pauli operators, including transversal logical Clifford gates or more complex dynamic programs.
Beyond QEC, the symbolic nature of this simulator may prove useful to design quantum circuits for variational quantum algorithms or error mitigation techniques. 

We also aim to improve simulation speed and scale via algorithmic and software optimisations. 
The simple and intuitive state representation reveals opportunities to reduce the number of terms required to compute a trace, including early identification and discard of terms containing single $O$ operators. 
Another opportunity is the factorisation of the tableau into products of non-correlated tableaux that can be processed independently, as shown in the example in Section~\ref{sec:example}. 
Sacrificing exactness, for example by sampling parts of the Pauli noise channels, may further accelerate computations with a controllable approximation error.
We plan to expand the symbolic capabilities to obtain truncated expressions for any quantity of interest while avoiding storing the full objects in memory.

Given its simplicity and many convenient features, especially for QEC, we expect SyQMA to develop rapidly into a powerful classical simulator that accelerates the advancement of quantum computation on several fronts.

\acknowledgements{}

We would like to thank the entire Quantinuum QEC team for their support. We also thank Dan Browne, Etienne Granet and Pablo Andr\'es-Mart\'inez for useful discussions, and Christopher Self and Cristina Cirstoiu for reviewing this manuscript.

\bibliographystyle{alpha}
\bibliography{ref}

\appendix

\section{Relation to PTM eigenvalues and the Walsh-Hadamard transform}
\label{app:wh_transform}

Let us remind ourselves that we seek to express $\mathcal{E}$ as a composition of Pauli-flip channels:
\begin{equation} \label{eq:channel_decomposition}
    \mathcal{E} = \bigcirc_{P \in \mathcal{P}_{\mathcal{Q}} \setminus \{I\}} \mathcal{E}^{p'_P}_P.
\end{equation}
To understand the origin of this decomposition, we examine the action of both sides of Eq.~\eqref{eq:channel_decomposition} on an arbitrary Pauli operator $R \in \mathcal{P}_{\mathcal{Q}}$.
On the left-hand side, since $PRP = (-1)^{\eta(P,R)} R$, the action of the multi-Pauli channel scales $R$ by a factor $\lambda_R$:
\begin{equation} \label{eq:lambda_def}
    \mathcal{E}(R) = \lambda_R \, R, \qquad \lambda_R = \sum_{P \in \mathcal{P}_{\mathcal{Q}}} p_P\, (-1)^{\eta(P,R)} = 1 - 2\sum_{P \in \mathcal{A}(R)} p_P.
\end{equation}
On the right-hand side, each individual flip channel $\mathcal{E}(p'_P, P)$ contributes a factor of $\epsilon_P = 1 - 2p'_P$ when $P$ anticommutes with $R$, and $1$ otherwise. The composed channel thus gives:
\begin{equation} \label{eq:eigenvalue_factorisation}
    \lambda_R = \prod_{P \in \mathcal{P}_{\mathcal{Q}} \setminus \{I\}} \epsilon_P^{\,\eta(P,R)} = \prod_{P \in \mathcal{A}(R)} \epsilon_P.
\end{equation}
Taking the product over $R \in \mathcal{A}(P)$ and separately over $R \in \mathcal{C}(P)$, all flip factors cancel pairwise except the one associated with $P$ itself, because exactly half of the $4^k$ operators anticommute with any given non-identity Pauli. This isolates $\epsilon_P$:
\begin{equation} \label{eq:epsilon_from_lambda}
    \epsilon_P = \left( \frac{\displaystyle\prod_{R \in \mathcal{A}(P)} \lambda_R}{\displaystyle\prod_{R \in \mathcal{C}(P)} \lambda_R} \right)^{2/4^{k}}.
\end{equation}
Since $p'_P = \frac{1 - \epsilon_P}{2}$, we obtain the final form for the flip-channel quasi-probabilities $p'_P$ in terms of the scaling factors (eigenvalues) $\lambda_R$:
\begin{equation} \label{eq:flip_factor_from_eigenvalues}
    p'_P = \frac{1}{2} - \frac{1}{2} \left( \frac{\displaystyle\prod_{R \in \mathcal{A}(P)} \lambda_R}{\displaystyle\prod_{R \in \mathcal{C}(P)} \lambda_R} \right)^{2/4^{k}}.
\end{equation}
Substituting $\lambda_R$ from Eq.~\eqref{eq:lambda_def} into Eq.~\eqref{eq:flip_factor_from_eigenvalues} recovers the disjoint probability expression in Eq.~\eqref{eq:flip_factor_from_probs}. 

The coefficients $\lambda_R \equiv \lambda(R)$ are precisely the eigenvalues of the Pauli transfer matrix (PTM) of the channel $\mathcal{E}$. In the PTM representation, the vectorised state $|{\rho}\rangle\!\rangle = \sum_P \Tr(P\rho)\, |{P}\rangle\!\rangle$ is expanded in the Pauli basis, and a Pauli channel acts diagonally: $\mathbf{E}\, |{P}\rangle\!\rangle = \lambda(P)\, |{P}\rangle\!\rangle$. 
The definition of $\lambda_R$ in Eq.~\eqref{eq:lambda_def} is recognised as the Walsh--Hadamard transform:
\begin{equation}
    \lambda(P) = \sum_{P' \in \mathcal{P}_{\mathcal{Q}}} p_{P'}\, (-1)^{\eta(P',\, P)},
\end{equation}
which relates the disjoint probability distribution $\{p_P\}$ to the PTM eigenvalues $\{\lambda(P)\}$, with the inverse transform recovering the probabilities:
\begin{equation}
    p_P = \frac{1}{4^{|\mathcal{Q}|}} \sum_{P' \in \mathcal{P}_{\mathcal{Q}}} \lambda(P')\, (-1)^{\eta(P,\, P')}.
\end{equation}
The factorisation in Eq.~\eqref{eq:eigenvalue_factorisation} is therefore the statement that the PTM eigenvalues of the composed flip channels multiply, since the PTM of a composition of channels is the product of the individual PTMs and all Pauli channels are simultaneously diagonal in the Pauli basis. As such, the eigenvalues $\{\lambda_R\}$ can be computed from the disjoint error probabilities of the original channel via a fast Walsh-Hadamard transform in $\mathcal{O}(4^k k)$ time, after which Eq.~\eqref{eq:flip_factor_from_eigenvalues} provides the factorised flip-channel parameters used by the simulator.

In our case, it is actually more convenient to use the eigenvalue representation in the internal tableau of the simulator, as the number of parameters stored for each $n$-qubit Pauli channel reduces from $4^n / 2$ (number of indices of anticommuting Pauli terms) to a single index of the Pauli eigenvalue representing the channel at any given point.

\section{Origin of the new definitions} \label{app:definitions}
This appendix provides detail on the origin of the definitions of the flip and rotation qubits, flip and COS operators, and the modified trace.

Consider any quantum program composed of qubit initialisation and trace-out, Clifford gates, Pauli projections, Pauli-flip channels and (generally non-Clifford) arbitrary-angle Pauli rotations.
We then replace any Pauli-flip channel $\mathcal{E}_P^p$ and Pauli rotation $R_P^\theta$ by the following equivalent circuits: \\
\begin{minipage}[c]{0.48\textwidth}
\begin{equation} \label{eq:flip_injection}
\begin{quantikz}
    \lstick{$\rho$} & \gate{P} & \rstick{$\mathcal{E}_P^p(\rho)$} \\
    \lstick{$\ket{+}$} & \ctrl{-1} & \rstick{$\tfrac{I + \epsilon Z}{2}$}
\end{quantikz}
\end{equation}
\end{minipage}
\hfill
\begin{minipage}[c]{0.48\textwidth}
\begin{equation} \label{eq:rotation_injection}
\begin{quantikz}
    \lstick{$\rho$} & \gate{P} & \rstick{$R_P^\theta \rho R_P^{-\theta}$} \\
    \lstick{$\ket{+}$} & \ctrl{-1} & \rstick{$\tfrac{I + \cos(\theta)Z+\sin(\theta)Y}{2}$}
\end{quantikz}
\end{equation}
\end{minipage}
Note that the projections are equivalent to applying $\mathcal{E}_X^p$ or $R_X^\theta$ before a simple projection $(I+Z)/2$.

We define the \textit{set of flip qubits} $\mathcal{F}$ as those auxiliary qubits prepared to inject Pauli-flip channels and the \textit{set of rotation qubits} $\mathcal{R}$ as those prepared to inject Pauli rotations.
Together with the computational qubits $\mathcal{Q}$ they form the set of all qubits as the union $\mathcal{N}=\mathcal{Q}\sqcup\mathcal{R}\sqcup\mathcal{F}$.

Note that, aside from the projection of the auxiliary qubits, every other operation defined in Section~\ref{sec:background} can be simulated efficiently via Clifford simulation with a stabiliser tableau.
To solve this we absorb those projections into a \textit{generalised trace} $\Tr$ that depends on the qubits it acts on:
\begin{enumerate}
    \item $\Tr_q\left(\rho\right) = \tr_q\left(\rho\right)$ for every computational qubit $q\in\mathcal{Q}$,
    \item $\Tr_f\left(\rho\right) = 2\tr_f\left(\rho \tfrac{I + \epsilon_f Z_f}{2}\right)$ for every flip qubit $f \in \mathcal{F}$,
    \item $\Tr_r\left(\rho\right) = 2\tr_r\left(\rho \tfrac{I + \cos(\theta_r)Z_r + \sin(\theta_r)Y_r}{2}\right)$ for every rotation qubit $r \in \mathcal{R}$.
\end{enumerate}
We add the qubit labels as subindices to the flip factors $\epsilon_f = 1-2p_f$ and the rotation angles $\theta_r$ to keep track of which parameter corresponds to each auxiliary qubit.

Under the generalised trace, the Pauli operators no longer have vanishing trace, so we give them a different name to avoid confusion and facilitate keeping track of the operations executed by the program and how they affect the state by a simple inspection of the tableau.
We then define the flip operators and the COS operators on their respective qubits and state their key properties as:
\begin{enumerate}
    \item \textit{flip operators}: $F_f = Z_f$ for every flip qubit $f \in \mathcal{F}$,
    \item \textit{COS operators}: $C_r = Z_r$, $O_r = X_r$, and $S_r = Y_r$ for every rotation qubit $r \in \mathcal{R}$,
    \item as Pauli operators they satisfy $F_f^2 = I_f$, $C_r^2 = O_r^2 = S_r^2 = I_r$, and $C_rO_r=iS_r$,
    \item their generalised trace is $\Tr_f(F_f) = 2 \epsilon_f$ and $\Tr_f(I_f) = 2$ for all flip qubits,
    \item and $\Tr_r(C_r) = 2\cos(\theta_r)$, $\Tr_r(O_r) = 0$, $\Tr_r(S_r) = 2\sin(\theta_r)$, $\Tr_r(I_r) = 2$ for all rotation qubits.
\end{enumerate}
We choose the definition of the generalised trace so that the trace of the identity is $\Tr(I) = 2$ for every qubit.

The trace $\tr$ is then replaced by the generalised trace $\Tr$, and we simulate the evolution of the joint state $\rho$ on all qubits in $\mathcal{N}$ with SyQMA's framework.
The (generally mixed and unnormalised) state on the computational qubit set $\mathcal{Q}$ is recovered by taking the trace over the rotation and flip qubits: $\rho|_\mathcal{Q} = \tr_\mathcal{R}\tr_\mathcal{F}\rho$.
With these definitions, the stabiliser tableau formalism can still be used under Pauli-flip channels and (generally non-Clifford) Pauli rotations.

\end{document}